\documentclass{article}

\usepackage{arxiv}

\usepackage[utf8]{inputenc}                   
\usepackage[T1]{fontenc}                      
\usepackage{hyperref}                         
\usepackage{url}                              
\usepackage{booktabs}                         
\usepackage{amsfonts}                         
\usepackage{nicefrac}                         
\usepackage{microtype}                        
\usepackage[capitalise,noabbrev]{cleveref}    
\usepackage{graphicx}
\usepackage{subcaption}
\usepackage[round]{natbib}
\usepackage{doi}
\usepackage[frozencache,cachedir=minted-cache]{minted}
\usepackage{amstext}
\usepackage{multirow}
\usepackage{tikz}
\usepackage[minted,breakable]{tcolorbox}

\makeatletter
\def\maxwidth{\ifdim\Gin@nat@width>\linewidth\linewidth\else\Gin@nat@width\fi}
\def\maxheight{\ifdim\Gin@nat@height>\textheight\textheight\else\Gin@nat@height\fi}
\makeatother
\setkeys{Gin}{width=\maxwidth, height=\maxheight, keepaspectratio}

\newcommand{\wrappedttt}[1]{%
  \tikz[baseline=(char.base)]\node[anchor=south west, draw, rectangle, rounded corners, inner sep=1pt, minimum size=4mm, text height=2mm](char){\small\texttt{#1}} ;}

\title{T5APR: Empowering Automated Program Repair across Languages through Checkpoint Ensemble}

\date{}

\author{ \href{https://orcid.org/0000-0001-6596-3658}{\includegraphics[scale=0.06]{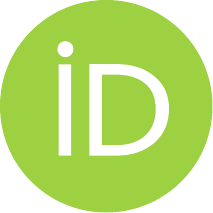}\hspace{1mm}Reza Gharibi}, Mohammad~Hadi Sadreddini, \href{https://orcid.org/0000-0002-9517-0541}{\includegraphics[scale=0.06]{images/orcid.pdf}\hspace{1mm}Seyed~Mostafa Fakhrahmad}\thanks{Corresponding author}\\
  Department of Computer Science and Engineering and IT \\
	School of Electrical and Computer Engineering \\
	Shiraz University, Shiraz, Iran \\
	\texttt{gharibi@cse.shirazu.ac.ir, sadredin@shirazu.ac.ir, fakhrahmad@shirazu.ac.ir} \\
}

\renewcommand{\undertitle}{}
\renewcommand{\shorttitle}{}

\hypersetup{
pdftitle={T5APR: Empowering Automated Program Repair across Languages through Checkpoint Ensemble},
pdfsubject={cs.SE, cs.LG},
pdfauthor={Reza Gharibi, Mohammad~Hadi Sadreddini, Seyed~Mostafa Fakhrahmad},
pdfkeywords={Automated program repair, Neural program repair, Deep learning, Transformer},
}

\begin{document}
\maketitle
\setcounter{footnote}{0}

\begin{abstract}
  Automated program repair (APR) using deep learning techniques has become an important area of research in recent years, aiming to automatically generate bug-fixing patches that can improve software reliability and maintainability. However, most existing methods either target a single language or require high computational resources to train multilingual models. In this paper, we propose T5APR, a novel neural program repair approach that provides a unified solution for bug fixing across multiple programming languages. T5APR leverages CodeT5, a powerful pre-trained text-to-text transformer model, and adopts a checkpoint ensemble strategy to improve patch recommendation. We conduct comprehensive evaluations on six well-known benchmarks in four programming languages (Java, Python, C, JavaScript), demonstrating T5APR's competitiveness against state-of-the-art techniques. T5APR correctly fixes 1,985 bugs, including 1,442 bugs that none of the compared techniques has fixed. We further support the effectiveness of our approach by conducting detailed analyses, such as comparing the correct patch ranking among different techniques. The findings of this study demonstrate the potential of T5APR for use in real-world applications and highlight the importance of multilingual approaches in the field of APR.
\end{abstract}


\section{Introduction}
\label{sec:introduction}

Software bugs are unavoidable in software development and can lead to security breaches, system failures, and user dissatisfaction, making it crucial to detect and fix them efficiently. However, manual debugging is time-consuming, particularly when dealing with large and intricate software systems. The growing need for high-quality and reliable software, coupled with the increasing complexity of software systems, has led to a surge of interest in automated program repair (APR). APR is an evolving research area that automatically fixes software bugs to enhance software reliability and maintainability. APR can potentially save developers time and effort, improve software quality and maintenance, and enable faster and more frequent software releases \citep{legouesAutomatedProgramRepair2019}.

Recent advancements in machine learning have shown promise in improving APR by adopting deep learning techniques, such as sequence-to-sequence, neural machine translation (NMT), and graph-to-sequence models, to automatically generate correct patches for buggy source code \citep{zhangSurveyLearningbasedAutomated2023, zhongNeuralProgramRepair2022}. These techniques can learn patterns from large code repositories and generate bug-fixing patches with state-of-the-art performance. APR tools that use these techniques are called neural program repair tools. Neural program repair has been mostly implemented with supervised learning on past bug-fixing commits to generate patches as sequences of tokens or edits, given a buggy code and its context \citep{chakrabortyCODITCodeEditing2022,chenSequenceRSequencetoSequenceLearning2019,dingPatchingTranslationData2020,jiangCURECodeAwareNeural2021,liDLFixContextbasedCode2020,lutellierCoCoNuTCombiningContextaware2020,yeNeuralProgramRepair2022,zhuSyntaxguidedEditDecoder2021}.

However, most existing APR methods are limited by their language-specificity and their high computational cost. They are either expensive to train for multiple programming languages \citep{lutellierCoCoNuTCombiningContextaware2020} or focus on a single language domain. Although they may generalize to other languages, they are rarely implemented and evaluated for other languages. This restricts their applicability and scalability across different languages and domains of code. For example, CoCoNuT \citep{lutellierCoCoNuTCombiningContextaware2020} and CIRCLE \citep{yuanCIRCLEContinualRepair2022} are two of the few approaches that are evaluated on multiple programming languages. To achieve multilingual repair, CoCoNuT trains separate models for each programming language, which requires large amounts of resources. CIRCLE uses continual learning on a single model but still needs measures to prevent catastrophic forgetting of the model and also a re-repairing post-processing strategy to provide patches for different languages.

\begin{figure}
  \centering
  \includegraphics{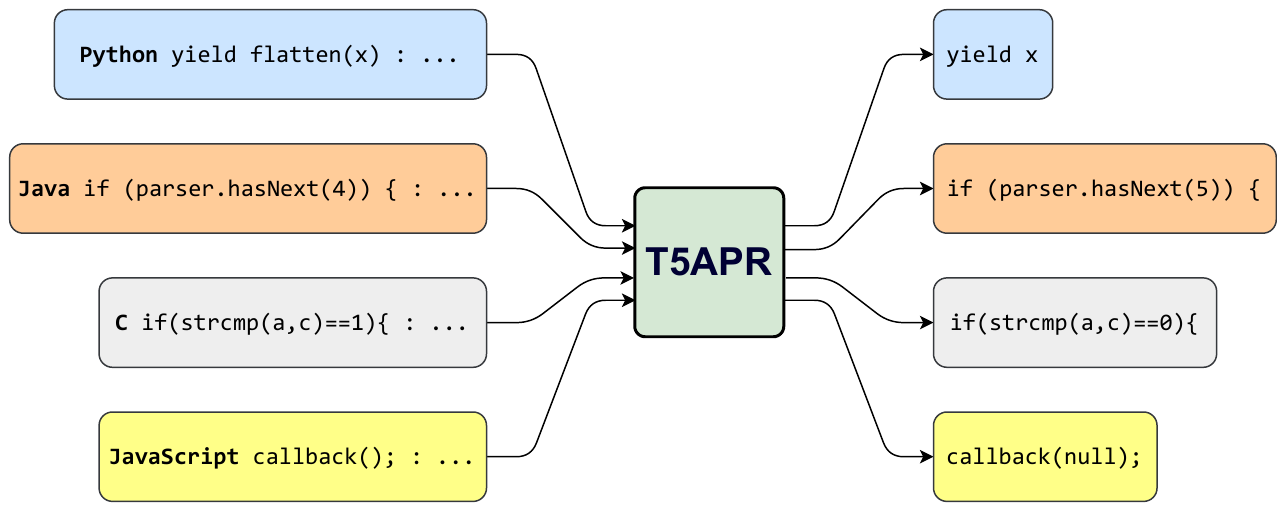}
  \caption{Illustration of T5APR for multilingual program repair.}
  \label{fig:t5apr}
\end{figure}

To address these limitations, this paper proposes T5APR, a novel multilingual neural repair method that leverages the power of transformer sequence-to-sequence models \citep{vaswaniAttentionAllYou2017}. T5APR is based on CodeT5 \citep{wangCodeT5IdentifierawareUnified2021}, a pre-trained model for code generation and understanding. T5APR fine-tunes CodeT5 in multitask learning style on a dataset of buggy and fixed code snippets and uses checkpoint ensemble \citep{chenCheckpointEnsemblesEnsemble2017} from different training steps to generate candidate patches. As shown in \cref{fig:t5apr}, we train a unified model to fix bugs from various programming languages and use language control codes in our input prompt to distinguish between different languages. This approach enables us to achieve efficient and scalable training with low resource consumption. T5APR then ranks and validates the generated candidate patches using the project's test suite to select the most suitable one.

We evaluate our method on six benchmarks including Defects4J \citep{justDefects4JDatabaseExisting2014}, Bears \citep{madeiralBEARSExtensibleJava2019}, QuixBugs \citep{linQuixBugsMultilingualProgram2017}, Codelfaws \citep{tanCodeflawsProgrammingCompetition2017}, ManyBugs \citep{legouesManyBugsIntroClassBenchmarks2015}, and BugAID \citep{hanamDiscoveringBugPatterns2016}, and compare its performance with 11 existing APR methods. Results show that T5APR can generate correct patches for various types of bugs in different languages and achieves state-of-the-art performance in terms of both repair effectiveness (i.e., correct fixes) and efficiency (i.e., ranking of fixes). Across all the benchmarks and from 5,257 bugs, T5APR fixes 1,985 of them, meaning the first plausible patch that it generates is semantically equivalent to the developer's patch.

The main contributions of this paper are as follows:

\begin{itemize}
  \item We introduce T5APR, a novel multilingual neural repair approach that offers a solution for bug fixing across various programming languages (\cref{sec:approach}).
  \item We address the challenge of resource efficiency in training without introducing additional models for each language by modeling multilingual APR with multitask learning (\cref{subsec:fine-tuning}).
  \item We propose a checkpoint ensemble strategy, enhancing the effectiveness of generated patches (\cref{subsec:checkpoint-ensemble}).
  \item We evaluate T5APR on six benchmarks and compare its performance with 11 state-of-the-art APR methods, achieving competitive performance (\cref{sec:experimental-setup,sec:results}).
  \item We provide an open-source implementation of T5APR, and our results are publicly available to foster further research and practical applications: \url{https://github.com/h4iku/T5APR}.
\end{itemize}

We discuss related work in \cref{sec:related-work}, and \cref{sec:conclusion} concludes the paper and suggests future directions.

\section{Approach}
\label{sec:approach}

\subsection{Overview}
\label{subsec:approach-overview}

\begin{figure}
  \centering
  \includegraphics{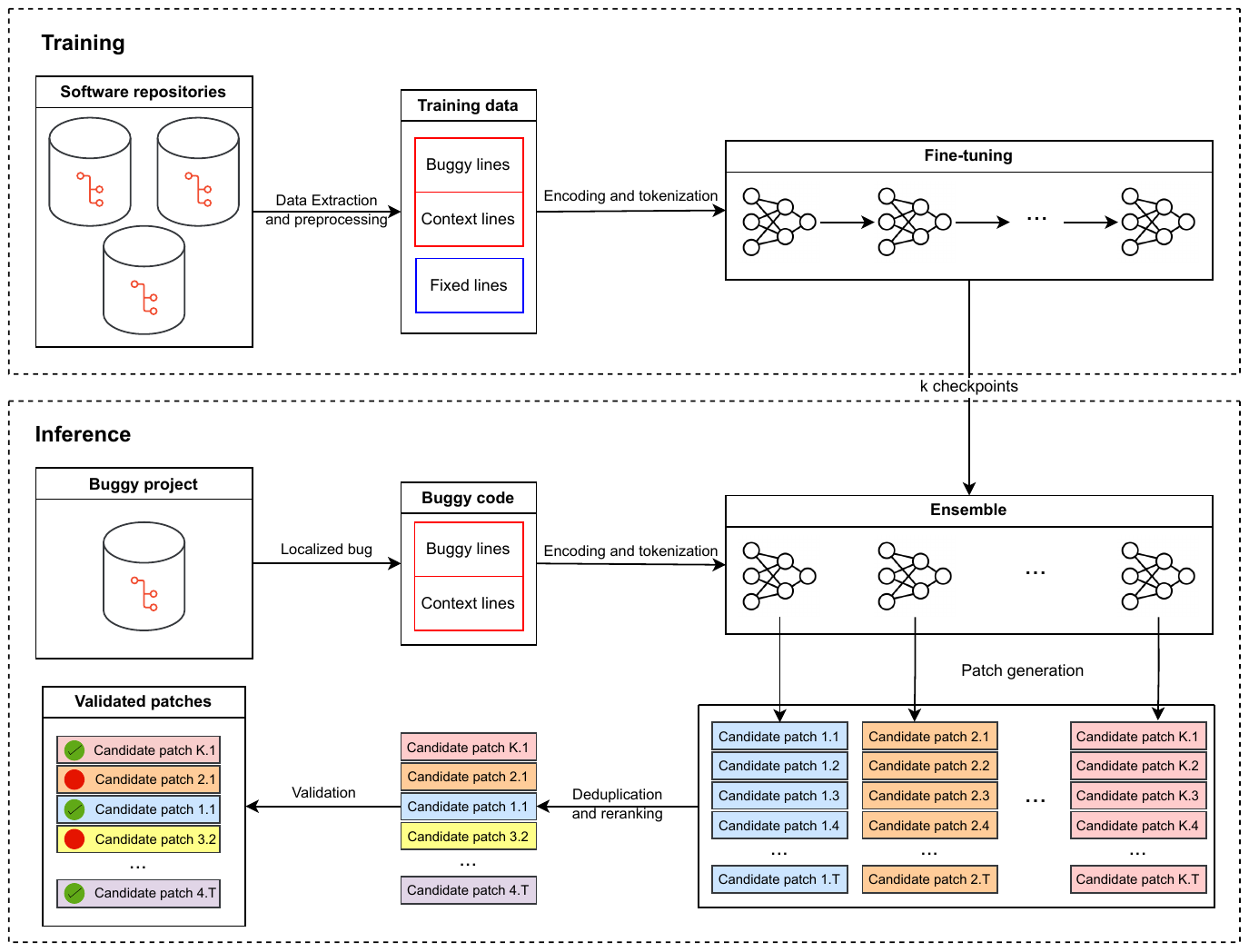}
  \caption{Overview of T5APR.}
  \label{fig:arch}
\end{figure}

\Cref{fig:arch} shows the overview of T5APR's structure. T5APR leverages CodeT5 as its base model and combines the outputs of multiple checkpoints to achieve improved performance in automated program repair (APR). T5APR involves two stages: training and inference. During the training stage, the CodeT5 model is fine-tuned on a large-scale multilingual dataset of buggy and fixed code snippets. In the inference stage, multiple checkpoints are used to generate candidate patches given buggy code and its context. The checkpoints are selected from different steps of the fine-tuning process. Finally, we rank the candidate patches based on a combination of their rank in each checkpoint and their likelihood score, and validate them using the project's test suite to select the most suitable patch. APR tools usually return the highest-ranked patch that compiles and passes the test cases, known as the first plausible patch. The following sections will describe this process in detail.

\subsection{Data extraction}
\label{subsec:data-extraction}

The process of training T5APR involves the extraction of vast amounts of data consisting of the buggy lines (source), their surrounding context to enhance the model's understanding of the faulty code, and their corresponding fixed versions (target). The training data is collected from multiple open-source projects in different programming languages, ensuring a broad representation of real-world code scenarios.

For the buggy context, we adopt the ``immediate buggy context,'' which refers to the function or method containing the buggy lines. Although other context choices, such as the entire file content or context obtained through data flow analysis or program slicing are possible, the immediate buggy context is often short and can easily be obtained for a multilingual model. The bigger the context, the more likely it is to have the needed ingredients to fix the bug \citep{yangWhereWereRepair2021}, but we have to give the model a longer sequence, which increases the computation time and resources. Therefore, we aim to create a balanced training environment that efficiently captures essential information for APR without sacrificing computational resources. The context-independence of our model allows for future incorporation of different context types to potentially improve performance \citep{chenSequenceRSequencetoSequenceLearning2019}.

In our experiments, we train the T5APR model on the dataset collected by CoCoNuT \citep{lutellierCoCoNuTCombiningContextaware2020}. This dataset follows the same criteria drawn above and consists of tuples of buggy, context, and fixed hunks of code extracted from the commit history of various open-source projects. A hunk is a set of consecutive lines of code change extracted from the commit history. For example, in \cref{lst:diff-flatten}, the lines that start with (\verb|-|) are buggy lines, lines that start with (\verb|+|) are fixed lines, and the whole \verb|flatten| function with the buggy lines is the context. The same applies to \cref{lst:diff-bears32}, except that the fixed hunk has more than one line here.

\begin{figure}
  \centering
  \begin{subfigure}[b]{0.45\textwidth}
    \begin{minted}[
      autogobble,
      fontsize=\small,
      frame=lines,
      framesep=2mm
      ]{diff}
      def flatten(arr):
          for x in arr:
              if isinstance(x, list):
                  for y in flatten(x):
                      yield y
              else:
    -             yield flatten(x)
    +             yield x 
    \end{minted}
    \caption{Code change of QuixBugs FLATTEN bug.}
    \label{lst:diff-flatten}
  \end{subfigure}%
  \hfill
  \begin{subfigure}[b]{0.45\textwidth}
    \begin{minted}[
      autogobble,
      fontsize=\small,
      frame=lines,
      framesep=2.22mm
      ]{diff}
      private boolean isFailOnCCE() {
    -     return getStep().isFailOnCCE();
    +     AbstractStep step = getStep();
    +     if (step == null) {
    +         return false;
    +     }
    +     return step.isFailOnCCE();
      }
    \end{minted}
    \caption{Code change of Bears-32 bug.}
    \label{lst:diff-bears32}
  \end{subfigure}
  \caption{Examples of buggy lines, buggy context, and fixed lines.}
\end{figure}

CoCoNuT uses a keyword-based heuristic to identify bug-fixing commits from their commit messages \citep{mockusIdentifyingReasonsSoftware2000}. They manually examined random commits and confirmed the effectiveness of this filtering process. However, we further clean the data during preprocessing to ensure high-quality training instances.

\subsection{Preprocessing}
\label{subsec:preprocessing}

This section describes the steps taken to prepare the input training data for the T5APR model. Before feeding the data to the model, we apply several preprocessing steps to enhance data quality and reduce computational resource requirements \citep{raffelExploringLimitsTransfer2020}:

\begin{itemize}
  \item \textbf{Comment removal}: Comments are removed from both sources and targets. This step ensures that the model focuses solely on the functional aspects of the code.
  \item \textbf{Deduplication}: We deduplicate the training data across source, context, and target based on their string representation, disregarding whitespace characters. This eliminates duplicate instances with the same functionality that only differ in whitespace characters, significantly reducing the dataset size without compromising the diversity of the code snippets.
  \item \textbf{Identical source and target removal}: Instances with identical source and target are discarded. This also includes instances with both empty source and target and instances where their source and target only differ in comments. Such instances do not represent actual bug fixes, and they provide no meaningful information for learning.
  \item \textbf{Empty target filtering}: Instances with empty targets are removed from the training data. Although this may negatively affect the model's ability to generate deletion operator patches, we demonstrate that the model can still effectively generate empty patches. In cases where the model does not produce an empty patch, because it is a single operator, we manually add an empty patch to the beginning of the patch list.
  \item \textbf{Source length filtering}: We filter instances based on the after-tokenization length of their source (excluding context). This step ensures that the code snippets are compatible with the model's input length constraints and only complete patches are used.
\end{itemize}

\subsection{Code representation and tokenization}
\label{subsec:data-prepare}

We represent buggy lines and their associated context in a unified format using a special delimiter token (\verb|:|) for the tokenization process:

\begin{verbatim}
input = prefix buggy_lines : context
\end{verbatim}

where \verb|prefix| is a language-specific control code that distinguishes different programming languages and is added to the beginning of each example, following previous works that use T5-based models \citep{berabiTFixLearningFix2021,raffelExploringLimitsTransfer2020,wangCodeT5IdentifierawareUnified2021}. In the case of multiline bugs, we concatenate lines using whitespace and put them right after each other. This input is then tokenized and truncated if necessary to ensure that its size remains below the model's maximum limit. We only use instances where their \verb|prefix + buggy_lines| and \verb|target_lines| lengths after tokenization are less than or equal to the model's maximum input and output sizes. Therefore, the truncation only affects the context part of each instance, if necessary.

We also tokenize the target in the same manner but independently and without adding a prefix or context. To tokenize inputs and targets, we use a pre-trained RoBERTa-based subword tokenizer, which uses byte-level byte-pair-encoding (BPE) \citep{sennrichNeuralMachineTranslation2016}. We use the tokenizer that comes with CodeT5 and is trained to be efficient in tokenizing source code. By using a tokenizer trained specifically on code, we can reduce the number of generated tokens, which in turn improves the model's training performance and output generation.

Subword tokenization algorithms split rare words into smaller, meaningful pieces while leaving common words intact. A BPE tokenizer allows us to include rare project-specific tokens in our vocabulary by breaking them into smaller pieces \citep{jiangCURECodeAwareNeural2021}. Subword tokenization gives the model a reasonable vocabulary size while trying to minimize the out-of-vocabulary (OOV) problem \citep{karampatsisBigCodeBig2020}.

The final output comprises tokenized source-context pairs and tokenized target labels. We represent the encoded input in the following format:

$\verb|<s>prefix |b_1 b_2 ... b_n \verb|:| c_1 c_2 ... c_m\verb|</s>|$

where $n$ and $m$ denote the number of buggy and context tokens, respectively. Tokens \verb|<s>| and \verb|</s>| mark the beginning and end of a sequence. \Cref{fig:encoded-tokens-examples} illustrates an example of how, given the source and context, the input is encoded for the considered programming languages.

\begin{figure}

  \hrule \vspace{2mm}

  \begin{subfigure}[b]{\textwidth}

    \textbf{Java}

    Source: \space \verb|return num % 2 != 0;|

    Context: \verb|public static boolean isEven(int num) { return num % 2 != 0; }|

    Encoded tokens:

    \foreach \word in {<s>, Java, Ġreturn, Ġnum, Ġ\%, Ġ2, Ġ!=, Ġ0, ;, Ġ:, Ġpublic, Ġstatic, Ġboolean, Ġis, Even, \string(, int, Ġnum, \string), Ġ\{, Ġreturn, Ġnum, Ġ\%, Ġ2, Ġ!=, Ġ0, ;, Ġ\}, </s>} {
        \wrappedttt{\word}
      }

  \end{subfigure}

  \vspace{2mm} \hrule \vspace{2mm}

  \begin{subfigure}[b]{\textwidth}

    \textbf{Python}

    Source: \space \verb|return num % 2 != 0|

    Context: \verb|def is_even(num): return num % 2 != 0|

    Encoded tokens:

    \foreach \word in {<s>, Python, Ġreturn, Ġnum, Ġ\%, Ġ2, Ġ!=, Ġ0, Ġ:, Ġdef, Ġis, \_, even, \string(, num, \string):, Ġreturn, Ġnum, Ġ\%, Ġ2, Ġ!=, Ġ0, </s>} {
        \wrappedttt{\word}
      }

  \end{subfigure}

  \vspace{2mm} \hrule \vspace{2mm}

  \begin{subfigure}[b]{\textwidth}

    \textbf{C}

    Source: \space \verb|return num % 2 != 0;|

    Context: \verb|int is_even(int num) { return num % 2 != 0; }|

    Encoded tokens:

    \foreach \word in {<s>, C, Ġreturn, Ġnum, Ġ\%, Ġ2, Ġ!=, Ġ0, ;, Ġ:, Ġint, Ġis, \_, even, \string(, int, Ġnum, \string), Ġ\{, Ġreturn, Ġnum, Ġ\%, Ġ2, Ġ!=, Ġ0, ;, Ġ\}, </s>} {
        \wrappedttt{\word}
      }

  \end{subfigure}

  \vspace{2mm} \hrule \vspace{2mm}

  \begin{subfigure}[b]{\textwidth}

    \textbf{JavaScript}

    Source: \space \verb|return num % 2 !== 0;|

    Context: \verb|function isEven(num) { return num % 2 !== 0; }|

    Encoded tokens:

    \foreach \word in {<s>, JavaScript, Ġreturn, Ġnum, Ġ\%, Ġ2, Ġ!==, Ġ0, ;, Ġ:, Ġfunction, Ġis, Even, \string(, num, \string), Ġ\{, Ġreturn, Ġnum, Ġ\%, Ġ2, Ġ!==, Ġ0, ;, Ġ\}, </s>} {
        \wrappedttt{\word}
      }

  \end{subfigure}

  \vspace{2mm} \hrule

  \caption{Examples of encoded input tokens of different programming languages.}
  \label{fig:encoded-tokens-examples}
\end{figure}

The \verb|Ġ| token is used to represent space character since this tokenizer has been trained to consider spaces as part of the tokens.

After tokenizing and preparing the data for each programming language, we concatenate data from different programming languages into a single dataset.

\subsection{Model architecture}
\label{subsec:base-model}

Our base model is CodeT5, a state-of-the-art transformer model that can handle both natural language and source code. CodeT5 uses the same encoder-decoder architecture as T5 \citep{raffelExploringLimitsTransfer2020} but is pre-trained on a large-scale dataset of code snippets and natural language descriptions from various programming languages. The authors used the CodeSearchNet dataset \citep{husainCodeSearchNetChallengeEvaluating2020} and another dataset they collected from BigQuery, which contains code snippets from eight programming languages (Ruby, JavaScript, Go, Python, Java, PHP, C, and C\#) and corresponding natural language descriptions extracted from public code repositories.

CodeT5's architecture consists of an encoder-decoder framework with multiple layers of self-attention mechanism. The encoder produces hidden representations of the input sequence, while the decoder uses them to generate the output sequence. The self-attention mechanism enables the model to focus on different parts of the input sequence and capture complex relationships between different parts of the source code \citep{wangCodeT5IdentifierawareUnified2021}.

We choose CodeT5 as our base model for several reasons. First, it is an open-source transformer model, pre-trained on multiple programming languages that is readily available for use and adaptation. Second, it has a robust and flexible encoder-decoder architecture that can handle different input and output formats and lengths. Third, it has a small version with only 60M parameters, which is computationally more efficient than other larger models but still achieves comparable results \citep{wangCodeT5IdentifierawareUnified2021}.

CodeT5 has been shown to achieve state-of-the-art performance on several code-related tasks, including code generation, code summarization, code refinement, and code completion. This means that CodeT5 has a strong foundation for handling code-related tasks and can potentially learn to perform program repair effectively \citep{jiangCURECodeAwareNeural2021}.

CodeT5 undergoes four pre-training tasks to acquire its code-aware capabilities:

\begin{enumerate}
  \item Masked span prediction that randomly selects spans of arbitrary lengths to mask and then uses the decoder to predict these masked spans marked with some sentinel tokens.
  \item Identifier tagging that trains the model to understand whether a code token is an identifier or not.
  \item Masked Identifier prediction that masks all identifiers in the code and uses a unique mask token for all occurrences of one specific identifier.
  \item Bimodal dual generation that considers the generation of natural text from source code and source code from natural text (NL $\leftrightarrow$ PL).
\end{enumerate}

All these tasks are formulated as a sequence-to-sequence task. Tasks 1, 2, and 3 are part of identifier-aware denoising pre-training that enhances the model's understanding of code syntax, structure, and semantics. Task 4 is effective for conversions between text and code, like generating comments for a code or generating code snippets based on a description like GitHub Copilot.

\subsection{Fine-tuning CodeT5}
\label{subsec:fine-tuning}

Fine-tuning is the process of adapting a pre-trained model to a specific task using task-specific data. In our approach, we fine-tune the CodeT5 model for multilingual APR. This involves training a unified model on a multilingual dataset comprising multiple programming languages; in our case Java, Python, C, and JavaScript. Our training data contains three columns: the buggy hunk (source), the surrounding buggy context function, and the corresponding fixed hunk (target).

The fine-tuning process adjusts the parameters of the pre-trained CodeT5 model to better suit the specific task of APR by minimizing a cross-entropy loss function that measures the discrepancy between the model's predictions and the ground-truth target fixes. As the model encounters task-specific data, it continues to learn and update its parameters to improve performance on the repair task. We fine-tune all languages simultaneously in batches that contain samples from all programming languages while using a prefix to identify each language. This approach leverages multitask learning (i.e., considering repairing each language as a separate task), enlarging the dataset, and allowing knowledge transfer between bug-fixing tasks in different languages. By using a multilingual base model and fine-tuning it with data combined from different languages (\cref{subsec:data-prepare}), we facilitate multilingual learning, where the model learns from bugs and fixes across various programming languages. Notably, this strategy proves particularly effective for handling bugs that are common across all languages \citep{berabiTFixLearningFix2021}.

We fine-tune the model for a specified number of epochs denoted as $i$ while saving a checkpoint every $j$ step, resulting in $k$ checkpoints as shown in \cref{fig:arch}.

\subsection{Checkpoint ensemble}
\label{subsec:checkpoint-ensemble}

Because of the diverse nature of bugs and fixes, a single model with optimal parameters may not generalize well \citep{lutellierCoCoNuTCombiningContextaware2020}. Ensemble learning has been a popular technique in machine learning for enhancing model performance and robustness \citep{dietterichEnsembleMethodsMachine2000}. In the context of transformer models, ensemble learning involves training multiple instances of the same model with different initialization or hyperparameters and then combining their results to obtain a final output.

Prior approaches in APR have utilized ensemble models, which combine multiple distinct models to generate bug-fixing patches \citep{jiangCURECodeAwareNeural2021,jiangKNODDomainKnowledge2023,lutellierCoCoNuTCombiningContextaware2020}. While this method has shown effectiveness in fixing more bugs, it often entails a significant computational cost due to training multiple specialized models with different inputs, hyperparameters, and, in some cases, even distinct architectures.

In contrast, we adopt a checkpoint ensemble approach for T5APR, which not only improves performance but also reduces training overhead \citep{chenCheckpointEnsemblesEnsemble2017}. Instead of training separate models, we exploit the diverse capabilities of the model at different training steps by saving and utilizing multiple checkpoints. We save $k$ checkpoints during the model training process, where $k$ represents the number of checkpoints used in the ensemble. The saved checkpoints have complementary abilities to generate patches for different types of bugs and contribute to the quality of patch ranking.

\subsection{Patch generation and ranking}
\label{subsec:generation-ranking}

Having obtained $k$ checkpoints from the trained model, we now proceed with generating patches for each bug. We apply the same data preparation and tokenization steps to the localized buggy hunk and its context as described in \cref{subsec:data-prepare}, with the only difference being that we truncate all long instances to match the model size without discarding any of them.

In some cases, the buggy hunk is not within a function. Although our training data always contains functions, we do not discard these bugs. Instead, we let the context be empty and try to generate patches using only the buggy lines.

Next, we generate candidate patches for a given example using the model. To achieve this, we use beam search with a specific beam size $t$ on each checkpoint, resulting in the generation of $t$ best patches from each checkpoint based on the maximum likelihood estimation score of each sequence. In total, we obtain $k \times t$ patches through the checkpoint ensemble as shown in \cref{fig:arch}.

To consolidate the generated patches from different checkpoints, we combine, deduplicate, and rerank patches by applying the following steps:

\begin{enumerate}
  \item We normalize whitespace characters in the generated patches to ensure consistency.
  \item We merge and sort the patches for each hunk according to their checkpoint ranks, breaking ties using the sequence scores (i.e., the likelihood score of each sequence generated by each checkpoint).
  \item We remove patches that are identical to the buggy source, as they do not contribute to the repair process.
  \item We deduplicate the patches, keeping only the unique ones, and retain the first patch in the list in case of duplicates.
  \item Finally, to account for the possibility of removing buggy lines, we add an empty patch at the beginning of the list for any bug that lacks one.
\end{enumerate}

For single-hunk bugs, the generated list is ready for validation. However, for multi-hunk bugs (bugs that require changes in more than one code location), we undertake additional processing to reduce the search space of patches \citep{sahaHarnessingEvolutionMultiHunk2019}. Specifically, we focus on multi-hunk patches that exhibit the same changes across all hunks \citep{madeiralLargescaleStudyHumancloned2021}. We identify identical patches among those generated for all hunks of a bug and retain only the patches present in all hunks. The patches are then sorted based on the maximum sequence score among each hunk's patches. Consequently, we obtain a list of patches that can be applied to all hunks, significantly reducing the number of patches to validate.

These final candidate patches undergo further validation to select the correct fixes.

\subsection{Patch validation}
\label{subsec:patch-validation}

Test suite-based APR uses test suite as program correctness specification. In this stage, we validate the candidate patches obtained from the previous stage by applying them to the original source code, compiling the patched code, and running the developer-written test suite. The goal is to filter out patches that do not compile or fail to pass the test cases in the project's test suite.

To validate the candidate patches, we follow these steps:
We apply each patch to the buggy location of the source code by replacing the buggy lines with the corresponding fixed lines from the generated patch. We then compile the patched code and run the test suite. To make this process faster, if possible, we first run the bug-triggering test cases, and if all of them pass, we proceed to run the rest of the test cases that make the buggy version pass to avoid regression. We make sure to omit flaky tests from this process. The patched program is considered valid if it passes all the test cases that the buggy project passed and passes the triggering test cases that previously failed on the buggy project. This validation approach aligns with common practices in many APR studies \citep{lutellierCoCoNuTCombiningContextaware2020}.

The resulting patches that pass the validation process are referred to as plausible patches. A plausible patch is a patch that satisfies the test suite but may not necessarily fix the underlying bug. However, it is essential to compare these plausible patches to the ground truth (i.e., developer-written patches) to assess whether they correctly fix the bug or only overfit to the test cases \citep{qiAnalysisPatchPlausibility2015, smithCureWorseDisease2015}. A patch is correct if it passes the test suite and has the same or equivalent semantics as the developer's patch.

\section{Experimental setup}
\label{sec:experimental-setup}

\subsection{Research questions}

The research questions that we aim to answer in this paper are:

\begin{itemize}
  \item \textbf{RQ1 (Effectiveness and generalizability)}: How does T5APR compare with state-of-the-art APR methods in terms of repair effectiveness and generalizability?
  \item \textbf{RQ2 (Multiple plausible patches)}: How does the consideration of multiple plausible patches improve T5APR's repair effectiveness?
  \item \textbf{RQ3 (Ablation study)}: What is the impact of checkpoint ensemble on T5APR's performance?
  \item \textbf{RQ4 (Multilingual and monolingual)}: How does the effectiveness of T5APR's multilingual model compare with monolingual models for each programming language?
\end{itemize}

\subsection{Datasets}

\paragraph{Training data}

We use the same dataset provided by CoCoNuT \citep{lutellierCoCoNuTCombiningContextaware2020} on GitHub\footnote{\url{https://github.com/lin-tan/CoCoNut-Artifact}} for training the T5APR model. The dataset consists of tuples of buggy, context, and fixed hunks of code from the commit history of various open-source projects hosted on platforms such as GitHub, GitLab, and BitBucket. The dataset covers multiple programming languages, including Java, Python, C, and JavaScript, making it ideal for training a multilingual APR model.
We choose these languages because they are widely used in the software industry and cover different paradigms and syntaxes. Moreover, they have high popularity and availability of training data and evaluation benchmarks for program repair.
\Cref{tab:train-data} provides a summary of the dataset statistics, including the cutoff year of collected data, the number of projects, instances before preprocessing, and the number of instances after preprocessing and tokenization for each programming language.

CoCoNuT finds the date of the earliest bug in each evaluation benchmark and collects commits that were made before that date, and discards instances committed after that to avoid overlapping train and evaluation data. The cutoff year in \cref{tab:train-data} is the year that the data is collected until that year. Other works have also used this data \citep{jiangCURECodeAwareNeural2021,jiangKNODDomainKnowledge2023,yeNeuralProgramRepair2022,yuanCIRCLEContinualRepair2022}.

\begin{table}
  \caption{Summary of training data instances before and after preprocessing.}
  \centering
  \begin{tabular}{lcrrrr}
    \toprule
    Language   & Cutoff year & Projects & Instances & After preprocessing & After size filtering \\
    \midrule
    Java       & 2006        & 45,180   & 3,241,966 & 1,125,599           & 1,009,268            \\
    Python     & 2010        & 13,899   & 480,777   & 302,727             & 264,842              \\
    C          & 2005        & 12,577   & 2,735,506 & 671,119             & 586,893              \\
    JavaScript & 2010        & 10,163   & 2,254,253 & 544,860             & 463,027              \\
    \bottomrule
  \end{tabular}
  \label{tab:train-data}
\end{table}

\begin{figure}
  \centering
  \includegraphics{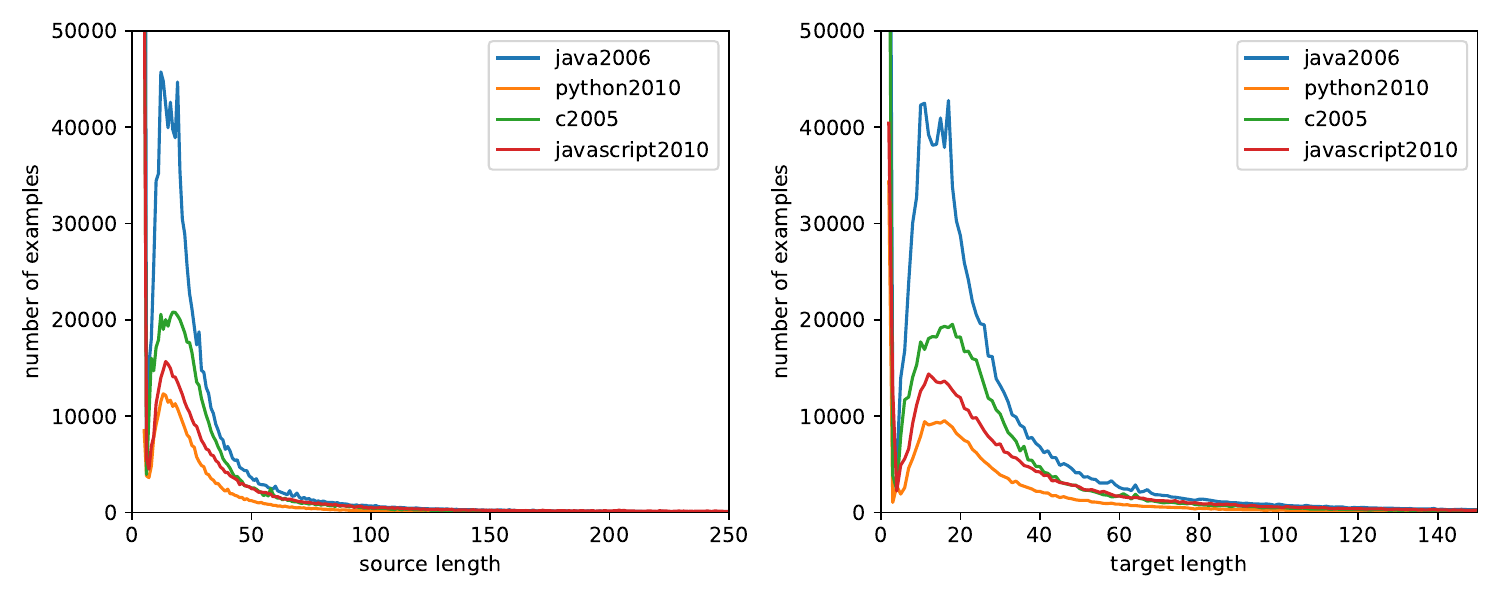}
  \caption{Distribution of training data instances by their token length.}
  \label{fig:length-stat}
\end{figure}

In our experiment, we use 512 tokens as the maximum input length of our approach, which is the same as the maximum input length of CodeT5. The maximum output length is set to 256 tokens to balance the computation cost and the model's performance. CodeT5 also uses 256 tokens as the maximum output length for its pre-training and some of its downstream tasks.
\Cref{fig:length-stat} shows the distribution of the source and target of training data instances based on their number of tokens after preprocessing but before size filtering. The $x$-axis indicates the token length range, while the $y$-axis represents the count of instances. Instances with shorter token lengths are more abundant, and as token length increases, the count of instances gradually decreases. This figure demonstrates that our choice of maximum input and output length is reasonable and covers most of the training data. \Cref{tab:train-data} shows that from the total of 2,644,305 instances after preprocessing, we retain 2,324,030 instances after size filtering, which is 87.89\% of the data.

\paragraph{Bug benchmarks}

We evaluate the performance of T5APR on a diverse set of benchmarks, spanning multiple programming languages and encompassing various types of bugs. We use the following benchmarks in our evaluation: Defects4J (Java) \citep{justDefects4JDatabaseExisting2014}, Bears (Java) \citep{madeiralBEARSExtensibleJava2019}, QuixBugs (Java and Python) \citep{linQuixBugsMultilingualProgram2017}, Codeflaws (C) \citep{tanCodeflawsProgrammingCompetition2017}, ManyBugs (C) \citep{legouesManyBugsIntroClassBenchmarks2015}, and BugAID (JavaScript) \citep{hanamDiscoveringBugPatterns2016}. These benchmarks collectively cover a wide range of real-world software defects and coding challenges \citep{sobreiraDissectionBugDataset2018,yeComprehensiveStudyAutomatic2021}.

Defects4J is a database and framework of real-world bugs from 17 well-known open-source Java projects. We follow prior work \citep{jiangKNODDomainKnowledge2023,yeNeuralProgramRepair2022,zhuSyntaxguidedEditDecoder2021} and separate Defects4J into two versions: Defects4J (v1.2) and Defects4J (v2.0). Defects4J (v1.2) contains 395 bugs, and Defects4J (v2.0) contains 444 additional bugs that are only available in the v2.0 version.
Bears benchmark is a collection of bugs from 72 Java projects hosted on GitHub and extracted using their continuous integration status history.
QuixBugs contains 40 bugs from the Quixey Challenge problems in both Java and Python. The programs are small classic algorithms in a single file.
Codeflaws is a set of bugs from Codeforces programming competition in C where each program is a single file.
ManyBugs contains bugs from large popular open-source C projects.
BugAID benchmark consists of 12 examples of common bug patterns in JavaScript described in \citet{hanamDiscoveringBugPatterns2016}.

\Cref{tab:benchmarks} provides detailed statistics for each benchmark. The table includes the number of bugs present in each benchmark, the count of bugs that are removed from consideration because they are either duplicates of other bugs or their buggy and fixed version has no change, the remaining number of bugs eligible for evaluation, and the total number of bugs that we attempt to repair. These statistics offer insights into the scale of each benchmark and the scope of our experimental evaluation.

\begin{table}
  \caption{Evaluation benchmark statistics.}
  \centering
  \begin{tabular}{lrrrr}
    \toprule
    Benchmark         & Bugs  & Removed & Remained & Attempted to Repair \\
    \midrule
    Defects4J (v1.2)  & 395   & 2       & 393      & 331                 \\
    Defects4J (v2.0)  & 444   & 0       & 444      & 357                 \\
    Bears             & 251   & 0       & 251      & 83                  \\
    QuixBugs (Java)   & 40    & 0       & 40       & 37                  \\
    QuixBugs (Python) & 40    & 0       & 40       & 40                  \\
    Codeflaws         & 3,903 & 7       & 3,896    & 3,863               \\
    ManyBugs          & 185   & 4       & 181      & 130                 \\
    BugAID            & 12    & 0       & 12       & 10                  \\
    \midrule
    Total             & 5,270 & 13      & 5,257    & 4,851               \\
    \bottomrule
  \end{tabular}
  \label{tab:benchmarks}
\end{table}

\subsection{Implementation details and parameters}

\paragraph{Implementation}
We implement T5APR in Python and use the Hugging Face Transformers library \citep{wolfTransformersStateoftheArtNatural2020} with PyTorch \citep{paszkePyTorchImperativeStyle2019} backend for training the model. Data preparation and preprocessing are performed using the Hugging Face Datasets library \citep{lhoestDatasetsCommunityLibrary2021}, which is based on Apache Arrow for efficient data processing.

We use the CodeT5 checkpoint that was trained using identifier-aware denoising pre-training objective for 100 epochs. CodeT5 has multiple variants with different sizes and number of parameters. We fine-tune the small model (\verb|CodeT5-small|) that has a total of 60M parameters. Although bigger models tend to perform better, it has been shown that the small model is also relatively capable \citep{wangCodeT5IdentifierawareUnified2021}. We leave using other model sizes of CodeT5 to future work due to resource limitations.

The CodeT5 tokenizer's vocabulary size is 32,100, of which 32,000 tokens are obtained from the pre-training dataset with non-printable characters and low-frequency tokens (occurring less than three times) filtered and 100 special tokens for padding (\verb|<pad>|), masking (\verb|<mask>|), marking the beginning and end of a sequence (\verb|<s>, </s>|), and representing unknown tokens (\verb|<unk>|).

The choice of hyperparameters, such as the learning rate, batch size, or number of training epochs can have a significant impact on the performance of the fine-tuned model. These hyperparameters are typically tuned using a separate validation set, which is held out from the training data and used to evaluate the model's performance on unseen examples.

We employ the Optuna optimization framework \citep{akibaOptunaNextgenerationHyperparameter2019} and use the AdamW optimizer \citep{loshchilovDecoupledWeightDecay2018} to conduct hyperparameter search. We randomly divide our Python training dataset into a training and a validation set, with 5,000 instances in the validation set and the rest in the training set. This separation is only for hyperparameter tuning, and later for training, we use the entire Python dataset.

The evaluation criteria for hyperparameter tuning are the exact match and the BLEU score \citep{papineniBleuMethodAutomatic2002}. Exact match is the ratio of instances where the predicted sequence exactly matches the ground truth sequence. BLEU score looks at how many n-grams in the model's output match the n-grams in the ground truth sequence and is a measure of the similarity between the model output and the ground truth sequence, which we compute using the sacreBLEU library \citep{postCallClarityReporting2018}. We define the objective metric for hyperparameter optimization as the sum exact match and BLEU score as follows:

\begin{equation}
  \text{objective metric} = \text{exact match} \times 100 + \text{BLEU score}
\end{equation}

The BLEU score from sacreBLEU ranges from 0 to 100, with 100 being the best possible score. Therefore, we multiply exact match by 100 to have the same range values for both metrics.

We define the hyperparameter search space based on the reasonable values that are commonly used for transformer fine-tuning in related work as follows: The learning rate ranges from $1e-5$ to $1e-3$, training epochs range from 1 to 5 epochs, training batch size has a search range from 4 to 16, beam size to 5, and learning rate scheduler type includes constant, cosine, linear, and polynomial.

After hyperparameter tuning, we set the final hyperparameters as follows: the train batch size is set to 8, the training epochs to 1, the learning rate to $1e-4$, and the learning rate scheduler type to constant. We also use mixed precision of FP16 for faster training. During training, we set $k=5$ and save five checkpoints. Each checkpoint is saved at every 20\% step of the training epoch.

In our decision to use five checkpoints, we draw inspiration from the previous successful approaches of CoCoNuT \citep{lutellierCoCoNuTCombiningContextaware2020}, CURE \citep{jiangCURECodeAwareNeural2021}, and KNOD \citep{jiangKNODDomainKnowledge2023}, which also employ five to ten models in their ensemble. This number has proven effective in related work, and we adopt it as a reasonable starting point for our ensemble. We show that more checkpoints result in more bug fixes, as also shown by CoCoNuT and CURE.

For the final inference and patch generation of benchmarks, we set the beam size to 100 to generate 100 patches from each checkpoint. A larger beam size would improve the results \citep{tufanoLearningMeaningfulCode2019}, but due to resource limitations, we chose a beam size of 100.

For parsing source files and extracting buggy context, we utilize the Tree-sitter\footnote{\url{https://tree-sitter.github.io/tree-sitter/}} parsing library and the lexers available in Pygments\footnote{\url{https://pygments.org/}}. These libraries can tokenize and parse many programming languages, making them suitable for our multilingual approach. We also use Unidiff\footnote{\url{https://github.com/matiasb/python-unidiff}} to parse the diff of the buggy and fixed codes to extract the location of buggy hunks.

\paragraph{Infrastructure}
We train our model on a server with 4 cores of an Intel Xeon Platinum 8259CL CPU, 16 GB RAM, and an NVIDIA T4 GPU with 16 GB VRAM. For evaluation, we use another system with a 6-core Intel Core i7-8750H CPU, 16 GB RAM, and an NVIDIA GeForce GTX 1060 GPU with 6 GB VRAM.

\subsection{Patch assessment}

Patches that can be compiled and pass the project's test suite are called plausible. However, a plausible patch may not fix the bug if the test suite is weak and does not cover all the cases \citep{qiAnalysisPatchPlausibility2015}. This is called the overfitting problem, where the patch only works for the test cases and not for the problem \citep{smithCureWorseDisease2015}. Therefore, we use the following criteria to determine if a plausible patch is correct \citep{yeITERIterativeNeural2024}:

\begin{itemize}
  \item It is identical to the developer-provided patch.
  \item It is identical to correct patches generated by existing techniques that have undergone public review by the community in open-source repositories.
  \item We judge it semantically equivalent to the developer-provided patch using rules described by \citet{liuEfficiencyTestSuite2020}.
\end{itemize}

To adhere to these criteria, one author checked whether the patches were identical to those created by the developer or other existing techniques. For the remaining patches that required semantic equivalence checking, the author consulted with another author in case of uncertainty.
To reduce the potential for errors in this process, we have made all generated patches publicly available for public judgment and review.\footnote{\url{https://github.com/h4iku/T5APR/tree/main/results}}

\subsection{Analysis procedure}

We compare T5APR against recent state-of-the-art learning-based tools and tools from other categories, such as template-based and semantic-based repair that are evaluated on our selected benchmarks and report their results under perfect fault localization setting. Techniques that use perfect fault localization are given the exact location of the bug. We identify the location of each bug with the help of human-written patches and their diff.
Some approaches use different fault localization algorithms or implementations to find the buggy location, which makes it difficult to only compare the repair capabilities of each approach. Recent studies suggest that perfect fault localization is the preferred way to evaluate APR approaches, as it allows a fair comparison of APR techniques without depending on the fault localization method \citep{liuYouCannotFix2019,liuEfficiencyTestSuite2020}.

We compare T5APR with 11 state-of-the-art tools, including seven Java APR tools: SequenceR \citep{chenSequenceRSequencetoSequenceLearning2019}, TBar \citep{liuTBarRevisitingTemplatebased2019}, DLFix \citep{liDLFixContextbasedCode2020}, CURE \citep{jiangCURECodeAwareNeural2021}, Recoder \citep{zhuSyntaxguidedEditDecoder2021}, RewardRepair \citep{yeNeuralProgramRepair2022}, and KNOD \citep{jiangKNODDomainKnowledge2023}. One C tool: SOSRepair \citep{afzalSOSRepairExpressiveSemantic2019}. Two tools that use large language models: Codex \citep{prennerCanOpenAICodex2022} and ChatGPT \citep{sobaniaAnalysisAutomaticBug2023}. Lastly, CoCoNuT \citep{lutellierCoCoNuTCombiningContextaware2020}, which is evaluated on all four programming languages.

We did not include CIRCLE \citep{yuanCIRCLEContinualRepair2022} in the evaluation since they have not validated their candidate patches using benchmarks' test suite and only reported exact match results across their generated patches.

To compare with these approaches, we follow the previous works and only consider the first plausible patch generated by T5APR that successfully compiles and passes the test suite \citep{durieuxEmpiricalReviewJava2019,liuTBarRevisitingTemplatebased2019,lutellierCoCoNuTCombiningContextaware2020}. There might be correct patches further down the plausible patch list, but we analyze those in another section.

We obtain results from each tool's paper or repository. The repository results, when available, are typically more recent and may contain corrections from the paper versions. For the tools that do not report results with perfect localization or for some benchmarks, we use results from other publications that evaluate these tools under perfect localization setting \citep{liuEfficiencyTestSuite2020,zhongStandUp4NPRStandardizingSetUp2023}.

To compute patch ranking, compilable patch rate, and unique bugs that T5APR can fix compared with other approaches, we obtain the list of candidate patches and fixed bugs for each approach on each benchmark from their respective repositories (Those that provide it).

\section{Results and discussion}
\label{sec:results}

\subsection{RQ1: Effectiveness and generalizability}

\Cref{tab:main-results} presents the results of evaluating the performance of T5APR on multiple benchmarks and against a selection of state-of-the-art APR tools. The table shows the name and the total number of considered bugs for each benchmark below its name.
The results are displayed as $c/p$ where $c$ is the number of correct patches that are ranked first as the first plausible patch by an APR technique, and $p$ is the total number of plausible patches. We also show, in parentheses, the number of bugs that have identical patches to the developer-written patch. A dash (-) indicates that the tool has not been evaluated on the benchmark or does not support the programming language of the benchmark to the best of our knowledge. For the ManyBugs and BugAID benchmarks, we could not validate their patches; therefore, we cannot show the number of plausible patches and only report the number of correct patches that we manually identified.

We highlight T5APR's performance on these benchmarks as follows.
Overall, T5APR fixes 1,985 bugs across all the benchmarks, with 1,413 of them patched identical to the developer's patch. Results show that T5APR outperforms or equals all other approaches on all the evaluated benchmarks except for Defects4J (v1.2) and ManyBugs.

For Defects4J, we observe that T5APR achieves competitive results, particularly in Defects4J (v2.0) where it generates correct fixes for 56 bugs and outperforms all other approaches. In Defects4J (v1.2), KNOD performs better than T5APR by fixing 71 bugs, while T5APR fixes 67 bugs. It should be noted that CoCoNuT, CURE, and KNOD use a substantially larger beam size of 1000 versus 100 that T5APR uses, and it has been shown that using larger beam size leads to more correct patches \citep{tufanoLearningMeaningfulCode2019}. Notice that if we consider all the generated plausible patches (\cref{tab:all-correct}), T5APR reaches 72 correct bugs. This shows the need for a better patch ranking strategy in future studies \citep{kangLanguageModelsCan2022}.

In the case of Bears, T5APR demonstrates its potential by correctly repairing 24 bugs, outperforming all the compared tools. Results in these two benchmarks indicate T5APR's capability in addressing real-world bugs in Java programs.

In the QuixBugs benchmark, which encompasses both Java and Python programs, T5APR shows robust performance, repairing 25 Java bugs and achieving correct repair for 29 Python bugs. The correct to plausible patch ratio for both versions is about 96\%, where only one of the generated plausible patches is not correct. In the QuixBugs (Java) version, T5APR fails to fix the SQRT bug, and in the QuixBugs (Python) version, it fails to fix the DEPTH\_FIRST\_SEARCH bug where it is correctly fixed further down in the patch list.

Similarly, in Codeflaws, a C programming language benchmark, T5APR showcases its robustness by achieving a substantial repair of 1,764 bugs, outperforming its only other contender, CoCoNuT. Turning to the ManyBugs benchmark, T5APR performance remains competitive by repairing 15 bugs, positioning itself among the top-performing tools but outperformed by SOSRepair. In the BugAID benchmark, T5APR achieves the highest repair rate, successfully fixing 5 out of the 12 bugs, demonstrating its competence in addressing JavaScript bugs.

Out of the bugs fixed by T5APR for Defects4J (v1.2), Defects4J (v2.0), Bears, Codeflaws, and ManyBugs benchmarks, 10, 4, 4, 36, and 2 of them are multi-hunk, respectively. These are fixed using the strategy described in \Cref{subsec:generation-ranking}. The remaining benchmarks either do not contain multi-hunk bugs or are not successfully fixed by T5APR.

Comparing T5APR against existing state-of-the-art methods, we consistently observe competitive or superior performance across various benchmarks. The overall results suggest T5APR's effectiveness in repairing a wide range of software defects and its effectiveness in handling bugs from different programming languages.

\begin{table}
  \caption{The number of correctly fixed bugs and comparison with state-of-the-art approaches. Results are from the first plausible patch by each method and are shown as \textit{correct/plausible (identical)}. Values in parentheses are bugs with identical patches to the developer's. (-) indicates data unavailability. The highest number of correct patches for each benchmark is highlighted in bold.}
  \centering
  \scriptsize
  \tabcolsep=0.08cm
  \begin{tabular}{lcccccccc}
    \toprule
    Tool                                                          & Defects4J (v1.2)    & Defects4J (v2.0)     & Bears               & QuixBugs (Java)     & QuixBugs (Python)   & Codeflaws                    & ManyBugs           & BugAID           \\

                                                                  & 393 bugs            & 444 bugs             & 251 bugs            & 40 bugs             & 40 bugs             & 3,896 bugs                   & 181 bugs           & 12 bugs          \\
    \midrule
    SOSRepair \citep{afzalSOSRepairExpressiveSemantic2019}        & -                   & -                    & -                   & -                   & -                   & -                            & \textbf{16/23} (2) & -                \\
    SequenceR \citep{chenSequenceRSequencetoSequenceLearning2019} & 12/19 (10)          & -                    & 16/26 (14)          & 15/16 (15)          & -                   & -                            & -                  & -                \\
    TBar \citep{liuTBarRevisitingTemplatebased2019}               & 53/84 (16)          & -                    & -                   & -                   & -                   & -                            & -                  & -                \\
    DLFix \citep{liDLFixContextbasedCode2020}                     & 39/68 (34)          & -                    & -                   & -                   & -                   & -                            & -                  & -                \\
    CoCoNuT \citep{lutellierCoCoNuTCombiningContextaware2020}     & 44/85 (26)          & 21/31 (16)           & 19/33 (16)          & 13/20 (13)          & 19/21 (15)          & 423/716 (255)                & 7/- (7)            & 3/- (3)          \\
    CURE \citep{jiangCURECodeAwareNeural2021}                     & 57/104 (37)         & 19/- (9)             & -                   & \textbf{25/34} (20) & -                   & -                            & -                  & -                \\
    Recoder \citep{zhuSyntaxguidedEditDecoder2021}                & 64/69 (46)          & -                    & 5/17 (1)            & 17/17 (-)           & -                   & -                            & -                  & -                \\
    RewardRepair \citep{yeNeuralProgramRepair2022}                & 45/- (38)           & 45/- (42)            & -                   & 20/- (20)           & -                   & -                            & -                  & -                \\
    Codex \citep{prennerCanOpenAICodex2022}                       & -                   & -                    & -                   & 14/- (-)            & 23/- (-)            & -                            & -                  & -                \\
    ChatGPT \citep{sobaniaAnalysisAutomaticBug2023}               & -                   & -                    & -                   & -                   & 19/- (13)           & -                            & -                  & -                \\
    KNOD \citep{jiangKNODDomainKnowledge2023}                     & \textbf{71/85} (49) & 50/82 (27)           & -                   & \textbf{25/30} (19) & -                   & -                            & -                  & -                \\
    \midrule
    T5APR                                                         & 67/94 (46)          & \textbf{56/103} (35) & \textbf{24/33} (12) & \textbf{25/26} (18) & \textbf{29/30} (24) & \textbf{1,764/2,359} (1,259) & 15/- (14)          & \textbf{5/-} (5) \\
    \midrule
    Total                                                         &                     &                      &                     & 1,985 (1,413)       &                     &                              &                    &                  \\
    \bottomrule
  \end{tabular}
  \label{tab:main-results}
\end{table}

\begin{figure}
  \centering
  \includegraphics{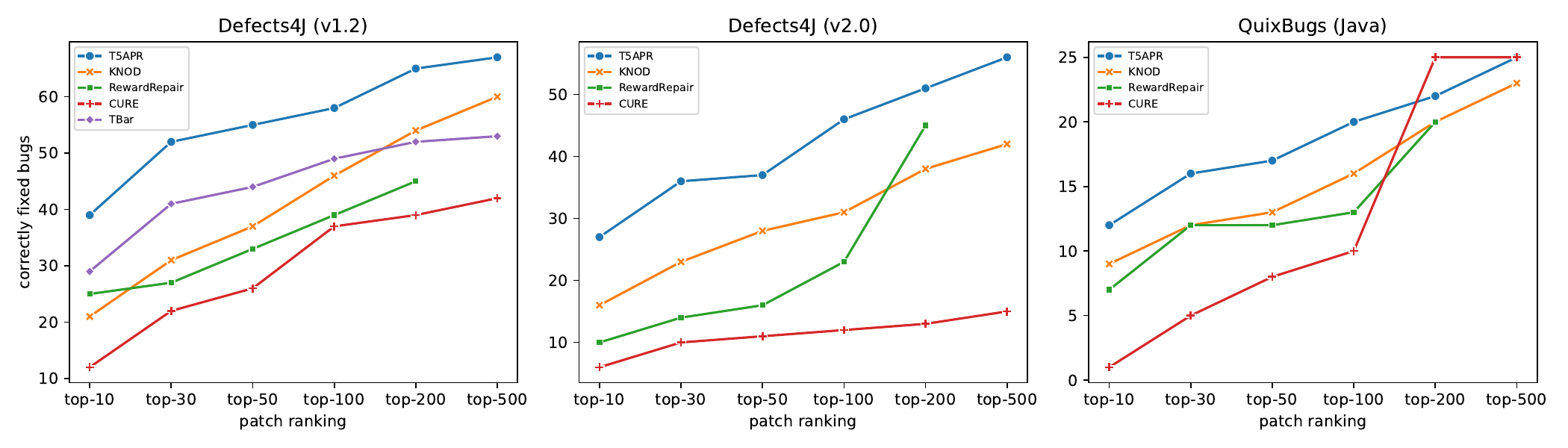}
  \caption{Ranking information of correct patches.}
  \label{fig:ranking-info}
\end{figure}

\paragraph{Patch ranking}
To provide a comprehensive understanding of the effectiveness of T5APR's patch ranking strategy, we analyze the ranking position of correct patches generated by each approach. We extract the ranking position of correct patches for each bug from the list of generated patches of each approach provided in their software repositories. Then, we calculate the number of bugs that are correctly fixed at various thresholds.
\Cref{fig:ranking-info} presents the distribution of correct patch ranking position information at different thresholds. Each line in the plot corresponds to a different approach, including T5APR, CURE, RewardRepair, and KNOD. The $x$-axis represents the ranking thresholds, while the $y$-axis indicates the number of correctly fixed bugs in each threshold. We only consider tools that we have access to their patch ranking information and generated candidate patches.

T5APR outperforms all other approaches in correct patch ranking except for top-200 on QuixBugs (Java), where CURE performs better. We can also see that although KNOD reaches better results than T5APR for Defects4J (v1.2) (71 vs. 67 bugs in \cref{tab:main-results}), T5APR fixes more bugs up to the top-500 generated candidate patches. KNOD generates fixes for the rest of the bugs in ranks higher than 500 due to using a larger beam size. Overall, 310 of the correct patches generated by T5APR are ranked first in the candidate patch list.

\begin{figure}
  \captionsetup[subfigure]{size=scriptsize, labelformat=empty}
  \centering
  \begin{subfigure}[b]{.25\textwidth}
    \centering
    \includegraphics[width=\linewidth]{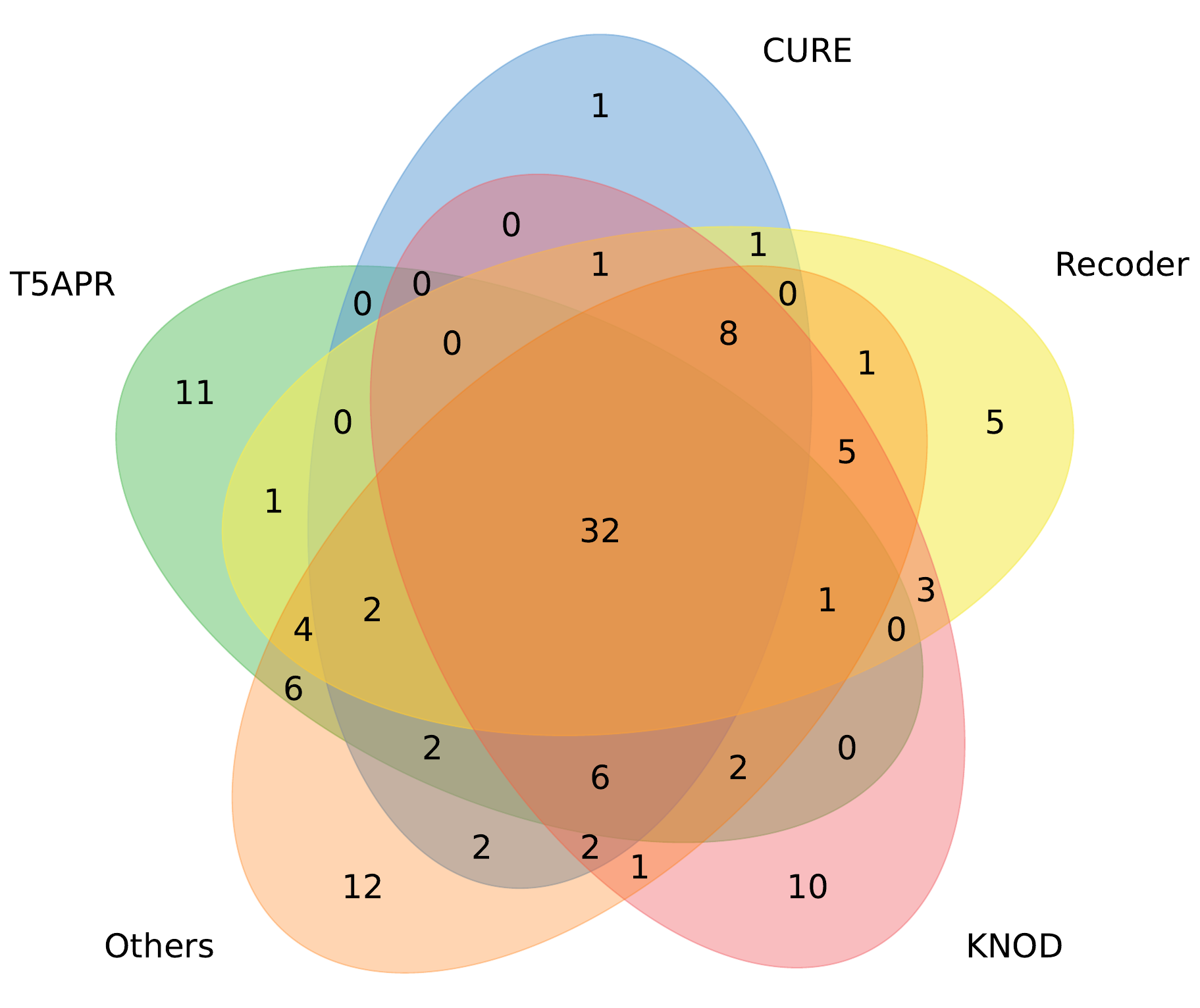}
    \caption{Defects4J (v1.2)}
  \end{subfigure}%
  \begin{subfigure}[b]{.25\textwidth}
    \centering
    \includegraphics[width=\linewidth]{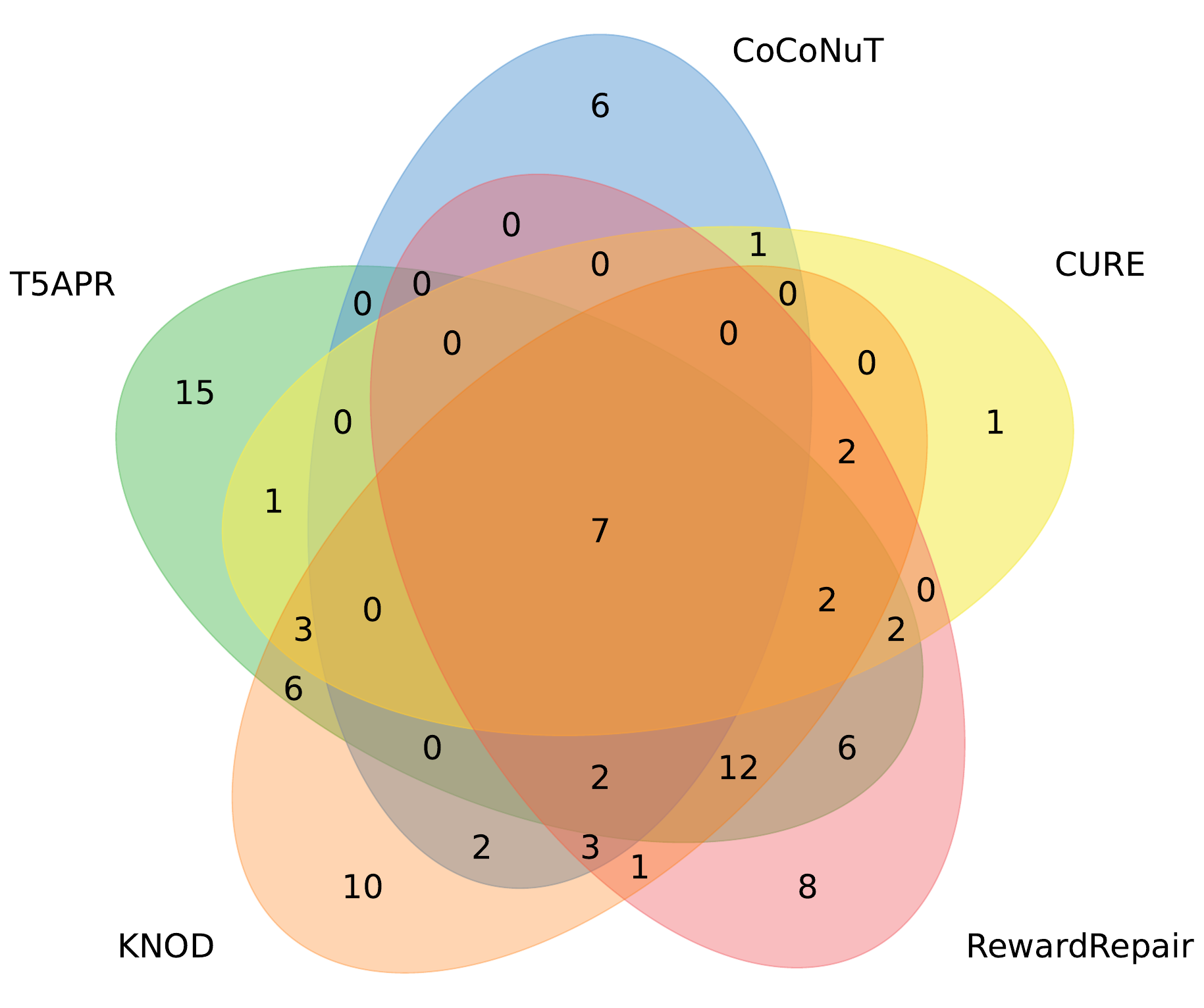}
    \caption{Defects4J (v2.0)}
  \end{subfigure}%
  \begin{subfigure}[b]{.25\textwidth}
    \centering
    \includegraphics[width=\linewidth]{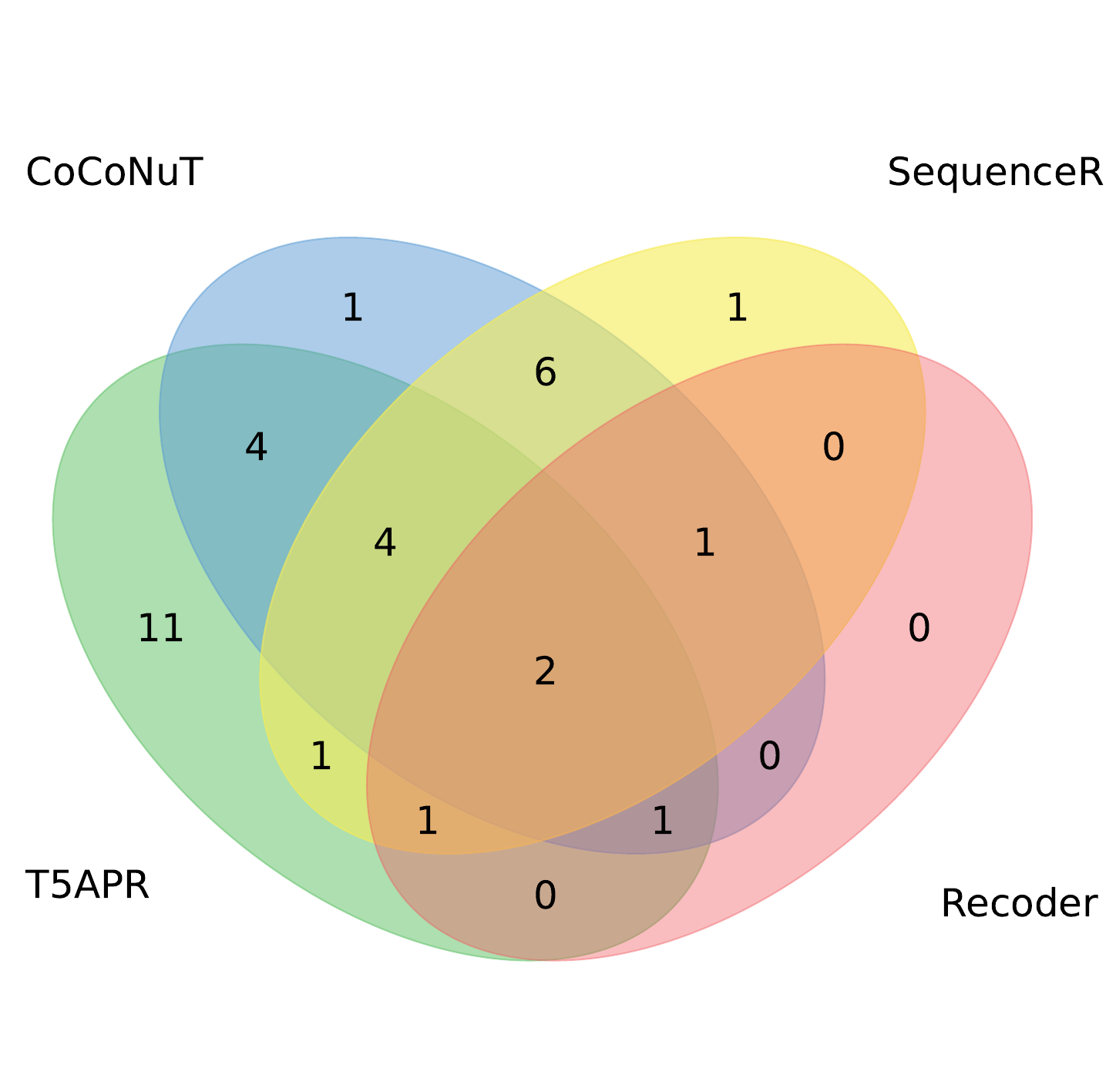}
    \caption{Bears}
  \end{subfigure}%
  \begin{subfigure}[b]{.25\textwidth}
    \centering
    \includegraphics[width=\linewidth]{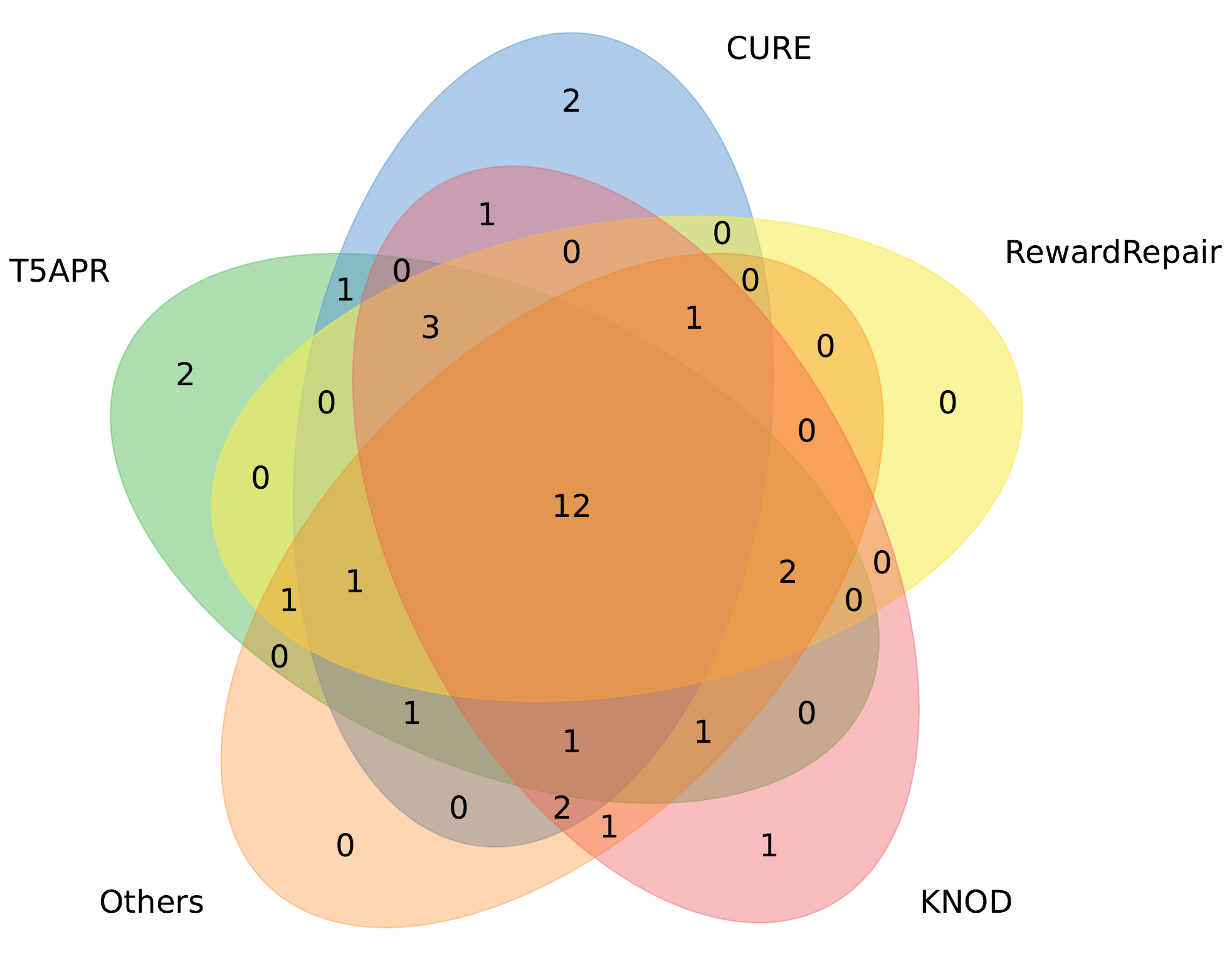}
    \caption{QuixBugs (Java)}
  \end{subfigure}
  \begin{subfigure}[b]{.25\textwidth}
    \centering
    \includegraphics[width=\linewidth]{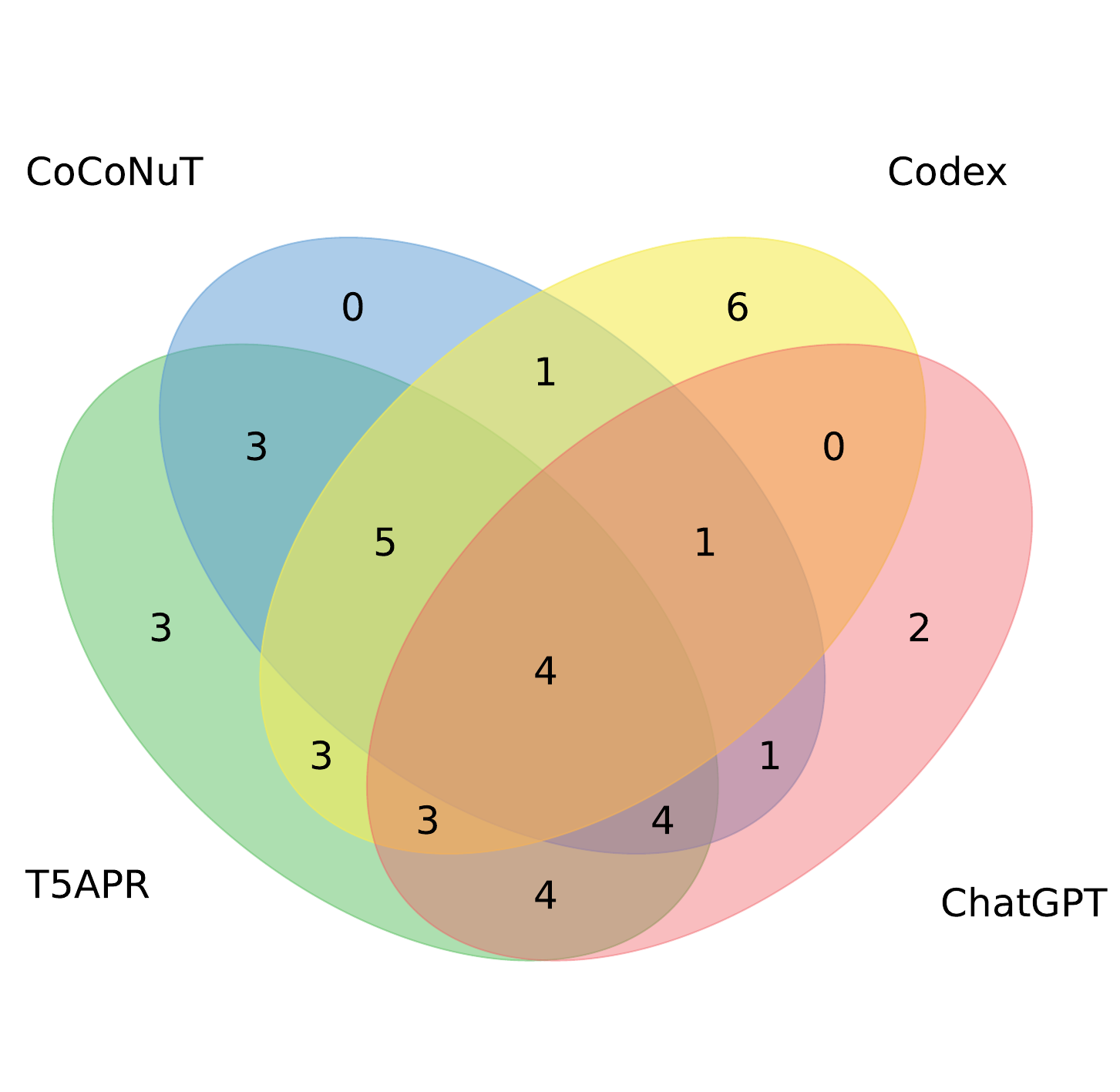}
    \caption{QuixBugs (Python)}
  \end{subfigure}%
  \begin{subfigure}[b]{.25\textwidth}
    \centering
    \includegraphics[width=\linewidth]{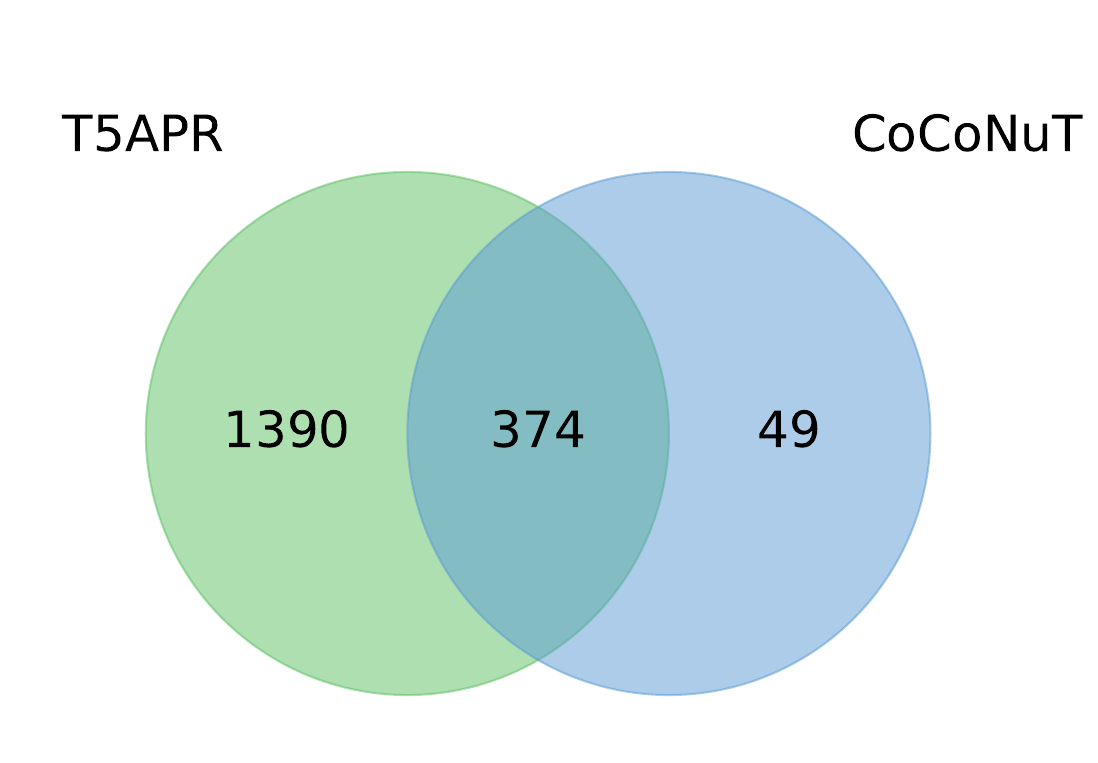}
    \caption{Codeflaws}
  \end{subfigure}%
  \begin{subfigure}[b]{.25\textwidth}
    \centering
    \includegraphics[width=\linewidth]{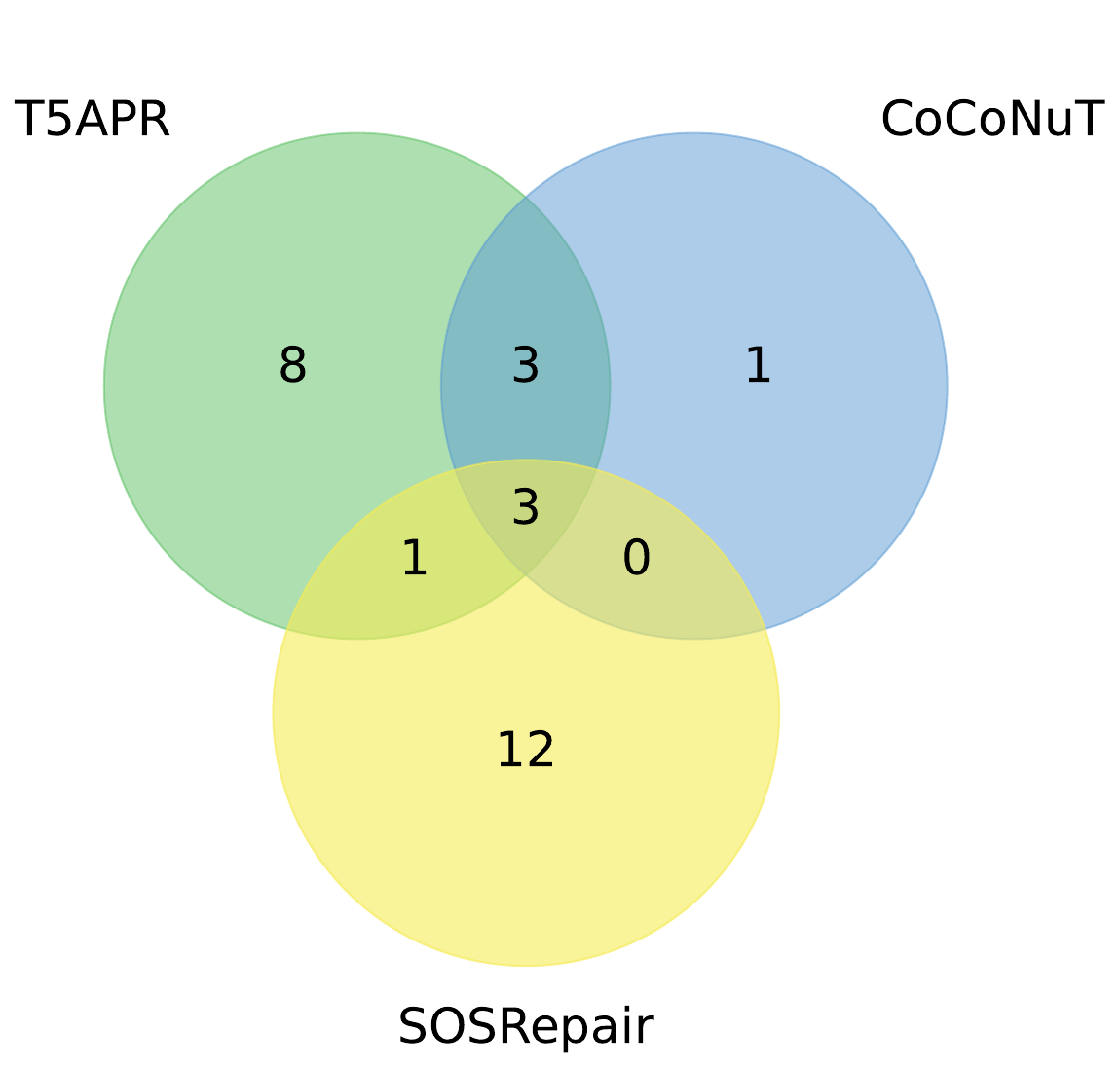}
    \caption{ManyBugs}
  \end{subfigure}%
  \begin{subfigure}[b]{.25\textwidth}
    \centering
    \includegraphics[width=\linewidth]{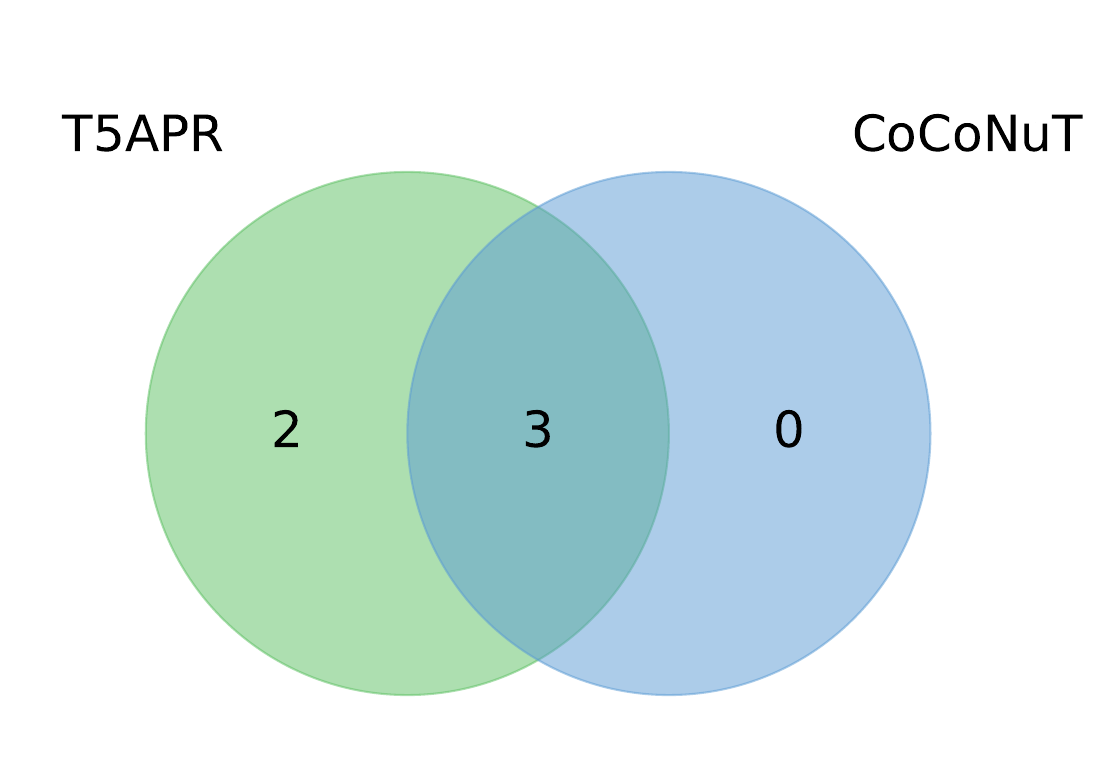}
    \caption{BugAID}
  \end{subfigure}
  \caption{Number of unique and overlapped bug fixes of T5APR and other tools.}
  \label{fig:unique-fixes}
\end{figure}

\paragraph{Unique bug fixes}
\Cref{fig:unique-fixes} presents the unique and overlapped number of bugs repaired by individual approaches from \cref{tab:main-results} for all the benchmarks. For benchmarks with more than four tools, we select the three best tools for that benchmark and combine the fixed bugs of the remaining tools under ``Others''. Across all the benchmarks, T5APR fixes 1,442 bugs that other tools do not fix. On the Defects4J benchmark, T5APR and KNOD complement each other by fixing 21 and 25 unique bugs for v1.2 and v2.0, respectively. On the ManyBugs benchmark, T5APR has good complementary quality and together with SOSRepair fixes 20 unique bugs. Overall, results show that T5APR complements compared existing works on all the evaluated benchmarks.

We provide a few examples of the bugs that T5APR can fix. \Cref{lst:fix-cli-40} shows the fix generated by T5APR and the ground-truth patch for Codec 5 bug from Defects4J (v2.0) benchmark that other tools do not fix. To fix this bug, T5APR notices that the header of the context method has \verb|throws ParseException|, which prompts T5APR to synthesize an exception throw statement. The only difference between T5APR and the developer's patch is the exception message, where T5APR uses the string parameter of the method while the developer writes a custom message.

\begin{figure}
  \centering
  \begin{subfigure}[b]{\textwidth}
    \begin{minted}[
      mathescape,
      autogobble,
      fontsize=\small,
      frame=lines,
      framesep=2mm,
      escapeinside=//
      ]{diff}
        public static <T> T createValue(final String str, final Class<T> clazz) throws ParseException {
            /\vdots/
      -     return null;
      +     throw new ParseException(str);
    \end{minted}
    \caption{T5APR's patch.}
  \end{subfigure}%
  \vspace{5mm}
  \begin{subfigure}[b]{\textwidth}
    \begin{minted}[
      mathescape,
      autogobble,
      fontsize=\small,
      frame=lines,
      framesep=2mm,
      escapeinside=//
      ]{diff}
      -     return null;
      +     throw new ParseException("Unable to handle the class: " + clazz);
    \end{minted}
    \caption{Developer-written patch.}
  \end{subfigure}
  \caption{Fix for Cli 40 bug from Defects4J (v2.0) benchmark.}
  \label{lst:fix-cli-40}
\end{figure}

\Cref{lst:fix-bears-32} gives the T5APR patch for Bears-32, which is only fixed by T5APR. For this bug, T5APR adds a null check for the returned value of \verb|getStep()|, which is a common pattern in Java. The developer's patch for this bug is shown in \cref{lst:diff-bears32}. T5APR's patch is semantically equivalent to the developer's patch.
\Cref{lst:fix-bears-46} shows another Bears benchmark bug patched by T5APR. This is a complete generation patch for Bears-46 with no lines to remove and is identical to the developer's patch. For this bug, T5APR also generates a null check and then based on the \verb|Set| return type of the context method, returns an empty set collection.
\Cref{lst:fix-quixbugs-lis} shows the patch for the same bug LIS from QuixBugs benchmark in both Java (\cref{lst:fix-quixbugs-lis-java}) and Python (\cref{lst:fix-quixbugs-lis-python}). The generated patch is similar for both languages, but T5APR adapts the syntax for each programming language.
\Cref{lst:fix-codeflaws-465B} shows the T5APR and developer's patch for multi-hunk bug 465-B-bug-16282461-16282524 from Codeflaws with identical patches for both hunk locations. T5APR finds the fix lower in the context and copies it to the buggy location. It also adds an if condition before it to avoid changing the value if it is already true.

\begin{figure}
  \begin{minted}[
    autogobble,
    fontsize=\small,
    frame=lines,
    framesep=2mm,
    ]{diff}
    - return getStep().isFailOnCCE();
    + return getStep()!= null && getStep().isFailOnCCE();
  \end{minted}
  \caption{T5APR's fix for Bears-32 bug from Bears benchmark.}
  \label{lst:fix-bears-32}
\end{figure}

\begin{figure}
  \begin{minted}[
    autogobble,
    frame=lines,
    framesep=2mm
    ]{diff}
      public Set<String> getMetadataKeys() {
    +     if (metadata == null) {
    +         return Collections.EMPTY_SET;
    +     }
          return metadata.keySet();
      } 
  \end{minted}
  \caption{T5APR's fix for Bears-46 bug from Bears benchmark.}
  \label{lst:fix-bears-46}
\end{figure}

\begin{figure}
  \centering
  \begin{subfigure}[b]{.45\textwidth}
    \begin{minted}[
      mathescape,
      autogobble,
      fontsize=\scriptsize,
      frame=lines,
      framesep=2mm,
      ]{diff}
        if (length == longest || val < arr[ends.get(length+1)]) {
            ends.put(length+1, i);
      -     longest = length + 1;
      +     if (length == longest) { longest = length + 1; }
        } 
    \end{minted}
    \caption{T5APR's Java patch.}
    \label{lst:fix-quixbugs-lis-java}
  \end{subfigure}%
  \hfill
  \begin{subfigure}[b]{.45\textwidth}
    \begin{minted}[
      mathescape,
      autogobble,
      fontsize=\scriptsize,
      frame=lines,
      framesep=3.23mm,
      ]{diff}
        if length == longest or val < arr[ends[length + 1]]:
            ends[length + 1] = i
      -     longest = length + 1
      +     if length == longest: longest = length + 1
    \end{minted}
    \caption{T5APR's Python patch.}
    \label{lst:fix-quixbugs-lis-python}
  \end{subfigure}%
  \vspace{5mm}
  \begin{subfigure}[b]{.45\textwidth}
    \begin{minted}[
      mathescape,
      autogobble,
      fontsize=\scriptsize,
      frame=lines,
      framesep=2mm,
      ]{diff}
      -     longest = length + 1;
      +     longest = Math.max(longest,length + 1);
    \end{minted}
    \caption{Developer's Java patch.}
  \end{subfigure}%
  \hfill
  \begin{subfigure}[b]{.45\textwidth}
    \begin{minted}[
      mathescape,
      autogobble,
      fontsize=\scriptsize,
      frame=lines,
      framesep=2mm,
      ]{diff}
      -     longest = length + 1
      +     longest = max(longest, length + 1)
    \end{minted}
    \caption{Developer's Python patch.}
  \end{subfigure}
  \caption{Fix for LIS bug from QuixBugs benchmark.}
  \label{lst:fix-quixbugs-lis}
\end{figure}

\begin{figure}
  \centering
  \begin{subfigure}[b]{\textwidth}
    \begin{minted}[
    autogobble,
    fontsize=\small,
    frame=lines,
    framesep=2mm
    ]{diff}
      for(i = 1; i < letters; ++i) {                                                
        current_letter = atoi(strtok(NULL, " "));
        if(current_letter && !inside_letter && actions != 0) {
    +     if(!inside_letter) inside_letter = true;
          actions += 2;
        }
        else if(current_letter) {
    +     if(!inside_letter) inside_letter = true;
          ++actions;
        }
        else {
          inside_letter = false;
        }
      }
  \end{minted}
    \caption{T5APR's patch.}
  \end{subfigure}%
  \vspace{5mm}
  \begin{subfigure}[b]{\textwidth}
    \begin{minted}[
    autogobble,
    fontsize=\small,
    frame=lines,
    framesep=2mm
    ]{diff}
    -
    +     inside_letter = true; 
  \end{minted}
    \caption{Developer's patch.}
  \end{subfigure}
  \caption{Fix for 465-B-bug-16282461-16282524 from Codeflaws benchmark.}
  \label{lst:fix-codeflaws-465B}
\end{figure}

\paragraph{Compilable patch rate}
Furthermore, we evaluate the compilable patch rate of candidate patches generated by different approaches, which reflects the tool's ability to generate syntactically correct and developer-like code. \Cref{tab:cr-compare} presents the average compilation rate across different top-X values of generated candidate patches.
T5APR has a slightly lower compilation rate than RewardRepair for all the top-X values. RewardRepair has a semantic training step that rewards compilable patches in its model training through backpropagation, which greatly helps it generate more compilable candidate patches. All tools have a decreasing compilation rate as the X value increases, which means that tools generate more non-compilable patches as they generate more candidate patches \citep{yeNeuralProgramRepair2022}.
We use an interval for the top-200 of CoCoNuT and CURE since they only report compilation rates for top-100 and top-1000 values. Moreover, we filter our considered bugs to be in the same set of bugs as RewardRepair for this comparison since we target more bugs. Like previous approaches \citep{jiangCURECodeAwareNeural2021,yeNeuralProgramRepair2022}, we combine Defects4J (v1.2) and QuixBugs (Java) patches. We directly list the numbers reported by \citet{yeNeuralProgramRepair2022} for SequenceR, CoCoNuT, CURE, and RewardRepair.

\begin{table}
  \caption{Average compilable patch rate of the top-X candidate patches in Defects4J (v1.2) and QuixBugs (Java).}
  \centering
  \begin{tabular}{lccc}
    \toprule
    Model        & Top-30 & Top-100 & Top-200   \\
    \midrule
    SequenceR    & 33\%   & -       & -         \\
    CoCoNuT      & 24\%   & 15\%    & 6\%-15\%  \\
    CURE         & 39\%   & 28\%    & 14\%-28\% \\
    RewardRepair & 45.3\% & 37.5\%  & 33.1\%    \\
    \midrule
    T5APR        & 42.7\% & 36.2\%  & 32.3\%    \\
    \bottomrule
  \end{tabular}
  \label{tab:cr-compare}
\end{table}

\Cref{tab:compilation-rate} shows the average compilable patch rate of the top-X candidate patches for each benchmark. The Codeflaws benchmark has the highest compilation rate among all the benchmarks, followed by QuixBugs (Java). The Bears benchmark has the lowest compilation rate among all the benchmarks, followed by Defects4J (v2.0). This may indicate some characteristics of Codeflaws and QuixBugs (Java) bugs that make them easier to compile than other benchmarks and some characteristics of Bears and Defects4J (v2.0) bugs that make them harder to compile.

\begin{table}
  \caption{Average compilable patch rate of the top-X candidate patches.}
  \centering
  \begin{tabular}{lccc}
    \toprule
    Benchmark        & Top-30 & Top-100 & Top-200 \\
    \midrule
    Defects4J (v1.2) & 33.2\% & 30.1\%  & 27.7\%  \\
    Defects4J (v2.0) & 31.8\% & 28.8\%  & 26.3\%  \\
    Bears            & 34.6\% & 26.4\%  & 23.6\%  \\
    QuixBugs (Java)  & 49.1\% & 42.6\%  & 37.1\%  \\
    Codeflaws        & 73.5\% & 70.3\%  & 67.3\%  \\
    \midrule
    Overall          & 67.8\% & 64.8\%  & 61.9\%  \\
    \bottomrule
  \end{tabular}
  \label{tab:compilation-rate}
\end{table}

\paragraph{Validation time cost}

We also assess the time cost of the validation effort required to reach plausible and correct patches. \Cref{tab:val-cost-plausible,tab:val-cost-correct} provide an overview of the time, the number of validated patches, and the maximum instances of timeouts it takes until the first plausible and correct patch is found, respectively.
In both cases, Codeflaws has both the lowest and highest time to reach a plausible and correct patch because on the one hand, Codeflaws programs are small and fast to compile and test, but on the other hand they have the highest number of timeout patches between benchmarks. A large portion of the validation time for benchmarks is spent on bugs with timeout patches, as we need to wait for a fixed duration until each patch times out before proceeding.

The number of patch candidates (NPC) validated until the first plausible patch is found is a metric to measure the repair efficiency of an APR tool \citep{liuEfficiencyTestSuite2020}. We can see from \cref{tab:val-cost-plausible} that for all the benchmarks, the median NPC is lower than 15 candidate patches.
The overall time to find a plausible patch ranges from 0.299 seconds to about 3 hours, with a median of 6.182 seconds and a mean of 1 minute and 31.416 seconds. The overall time to find a correct patch ranges from 0.333 seconds to about 2.5 hours, with a median of 5.748 seconds and a mean of 1 minute and 13.520 seconds. This suggests that finding a correct patch is slightly faster than finding a plausible but incorrect patch on average. This finding supports the conclusion of \citet{liuEfficiencyTestSuite2020} that most of the time a tool reaches a correct patch faster than an incorrect but only plausible patch.

Overall, in our multilingual experiment, we validated 1,172,267 patches across all the benchmarks, which took about 27 days of execution time. Training the multilingual model for one epoch on a single GPU took about 17 hours.

\begin{table}
  \caption{Statistics for validation until a plausible patch. Time is in HH:MM:SS.fff format.}
  \centering
  \small
  \tabcolsep=0.15cm
  \begin{tabular}{lccccccccc}
    \toprule
    \multirow{2}{*}{Benchmark}                    &
    \multicolumn{4}{c}{Time to plausible}         &
    \multicolumn{4}{c}{Validated until plausible} &
    \multicolumn{1}{c}{Timeouts}                                                                                                                          \\
    \cmidrule(r){2-5} \cmidrule(r){6-9} \cmidrule(r){10-10}
                                                  & {min}        & {max}        & {median}     & {mean}       & {min} & {max} & {median} & {mean} & {max} \\
    \midrule
    Defects4J (v1.2)                              & 00:00:06.962 & 01:18:59.266 & 00:01:38.354 & 00:04:56.558 & 1     & 318   & 10.0     & 43.83  & 2     \\
    Defects4J (v2.0)                              & 00:00:04.437 & 00:34:36.672 & 00:01:16.711 & 00:03:09.811 & 1     & 322   & 15.0     & 57.88  & 6     \\
    Bears                                         & 00:00:08.577 & 01:24:41.593 & 00:03:13.275 & 00:09:24.018 & 1     & 224   & 9.0      & 33.12  & 0     \\
    QuixBugs (Java)                               & 00:00:07.328 & 01:15:08.906 & 00:01:17.977 & 00:06:55.860 & 1     & 250   & 14.0     & 60.65  & 67    \\
    QuixBugs (Python)                             & 00:00:00.630 & 00:20:09.557 & 00:00:03.852 & 00:01:24.590 & 2     & 269   & 7.5      & 57.53  & 20    \\
    Codeflaws                                     & 00:00:00.299 & 03:09:24.491 & 00:00:04.683 & 00:01:08.845 & 1     & 490   & 10.0     & 37.97  & 150   \\
    \midrule
    Overall                                       & 00:00:00.299 & 03:09:24.491 & 00:00:06.182 & 00:01:31.416 & 1     & 490   & 10.0     & 39.34  & 150   \\
    \bottomrule
  \end{tabular}
  \label{tab:val-cost-plausible}
\end{table}

\begin{table}
  \caption{Statistics for validation until a correct patch. Time is in HH:MM:SS.fff format.}
  \centering
  \small
  \tabcolsep=0.15cm
  \begin{tabular}{lccccccccc}
    \toprule
    \multirow{2}{*}{Benchmark}                  &
    \multicolumn{4}{c}{Time to correct}         &
    \multicolumn{4}{c}{Validated until correct} &
    \multicolumn{1}{c}{Timeouts}                                                                                                                        \\
    \cmidrule(r){2-5} \cmidrule(r){6-9} \cmidrule(r){10-10}
                                                & {min}        & {max}        & {median}     & {mean}       & {min} & {max} & {median} & {mean} & {max} \\
    \midrule
    Defects4J (v1.2)                            & 00:00:06.962 & 00:37:33.473 & 00:01:23.657 & 00:03:07.288 & 1     & 234   & 8.0      & 31.22  & 2     \\
    Defects4J (v2.0)                            & 00:00:04.437 & 00:18:58.074 & 00:01:00.709 & 00:02:54.237 & 1     & 322   & 12.0     & 53.66  & 3     \\
    Bears                                       & 00:00:20.088 & 01:24:41.593 & 00:03:21.152 & 00:10:33.795 & 1     & 194   & 8.5      & 32.42  & 0     \\
    QuixBugs (Java)                             & 00:00:07.328 & 00:29:58.495 & 00:00:48.338 & 00:04:12.138 & 1     & 250   & 14.0     & 56.64  & 13    \\
    QuixBugs (Python)                           & 00:00:00.630 & 00:20:09.557 & 00:00:02.718 & 00:01:27.222 & 2     & 269   & 7.0      & 58.76  & 20    \\
    Codeflaws                                   & 00:00:00.333 & 02:33:47.090 & 00:00:04.578 & 00:00:55.622 & 1     & 358   & 9.0      & 37.74  & 150   \\
    \midrule
    Overall                                     & 00:00:00.333 & 02:33:47.090 & 00:00:05.748 & 00:01:13.520 & 1     & 358   & 9.0      & 38.45  & 150   \\
    \bottomrule
  \end{tabular}
  \label{tab:val-cost-correct}
\end{table}

\begin{tcolorbox}[title=RQ1 takeaways]

  T5APR fixes 1,985 bugs across six benchmarks in various languages, 1,413 of them identically to the developer's patch. It ranks a high number of correct patches within the top positions of generated candidates, with 310 of them ranked first. It also fixes 1,442 unique bugs that other tools cannot, showing its complementarity to other tools. The compilable patch rate of T5APR remains within a reasonable range, and its validation cost varies with the benchmark and the number of timeout patches, with an overall median of 10 patches to reach a plausible one. The findings demonstrate T5APR's effectiveness, capability, and efficiency in repairing a wide range of bugs in different programming languages.

\end{tcolorbox}

\subsection{RQ2: Multiple plausible patches}

\Cref{tab:all-correct} compares the results when only the first plausible patch is considered versus when all plausible patches are considered. The ``top-X'' thresholds consider only the first X plausible patches generated. The ``all'' threshold encompasses all plausible patches generated for each bug.

We find 2,309 correct patches when we consider all plausible patches, which is an increase of 344 from the first plausible patch. This means that 344 correct patches are not ranked as the first plausible patch by T5APR, and an incorrect patch passes the test cases due to test suite limitation. This limitation is an issue that most test suite-based APR tools have in common. The number of correct patches increases the most when we raise the threshold from top-1 to top-5, which means that most of the correct patches are ranked within the top-5 plausible patches by T5APR. This is important since according to a recent study, 72\% of developers are only willing to review up to five patches in practice \citep{nollerTrustEnhancementIssues2022}. The increase in the number of correct patches is smaller when the threshold is raised from top-5 to all.

\begin{table}
  \caption{Number of correctly fixed bugs based on plausible ranking when all plausible patches are considered.}
  \centering
  \begin{tabular}{lcccc}
    \toprule
    Benchmark         & Top-1 & Top-5 & Top-10 & All   \\
    \midrule
    Defects4J (v1.2)  & 67    & 72    & 72     & 72    \\
    Defects4J (v2.0)  & 56    & 64    & 65     & 65    \\
    Bears             & 24    & 25    & 25     & 26    \\
    QuixBugs (Java)   & 25    & 25    & 25     & 25    \\
    QuixBugs (Python) & 29    & 30    & 30     & 30    \\
    Codeflaws         & 1,764 & 1,990 & 2,017  & 2,071 \\
    ManyBugs          & -     & -     & -      & 15    \\
    BugAID            & -     & -     & -      & 5     \\
    \midrule
    Total             & 1,965 & 2,206 & 2,234  & 2,309 \\
    \bottomrule
  \end{tabular}
  \label{tab:all-correct}
\end{table}

\begin{tcolorbox}[title=RQ2 takeaways]

  By considering multiple plausible patches instead of only the first one, the repair effectiveness of T5APR improves by 17.5\%, increasing the number of fixed bugs from 1,965 to 2,309. Most of the correct patches are ranked within T5APR's top-5 plausible patches.

\end{tcolorbox}

\subsection{RQ3: Ablation study}

We employ an ensemble approach, combining the outputs of individual checkpoints to enhance the overall performance of T5APR. \Cref{tab:checkpoints-results} shows the performance of each checkpoint independently for different benchmarks. Each checkpoint result includes the manually added deletion patch. Compared with \cref{tab:main-results}, we can see that the combination of checkpoints has better results than each checkpoint independently. This demonstrates the effectiveness of our ensemble approach \citep{dietterichEnsembleMethodsMachine2000}. Overall, checkpoint5 fixes more bugs than others, but considering each benchmark independently, we can see that the best-performing checkpoint varies. This suggests that checkpoints complement each other and learn different patterns for different bugs.

\begin{table}
  \caption{Result of each checkpoint independently shown as \textit{correct/plausible}. Includes the manually added patch to each checkpoint.}
  \centering
  \begin{tabular}{lccccc}
    \toprule
    Benchmark         & checkpoint1 & checkpoint2 & checkpoint3 & checkpoint4 & checkpoint5 \\
    \midrule
    Defects4J (v1.2)  & 51/73       & 55/75       & 56/78       & 53/78       & 59/78       \\
    Defects4J (v2.0)  & 43/76       & 45/78       & 51/85       & 49/84       & 50/83       \\
    Bears             & 14/27       & 15/26       & 20/32       & 19/29       & 18/27       \\
    QuixBugs (Java)   & 15/17       & 15/16       & 16/17       & 18/18       & 18/18       \\
    QuixBugs (Python) & 18/19       & 19/20       & 22/23       & 23/24       & 24/24       \\
    Codeflaws         & 1,317/1,981 & 1,318/1,959 & 1,374/2,012 & 1,381/2,028 & 1,379/2,011 \\
    ManyBugs          & 11/-        & 12/-        & 12/-        & 13/-        & 12/-        \\
    BugAID            & 5/-         & 5/-         & 5/-         & 5/-         & 5/-         \\
    \midrule
    Total             & 1,474/2,193 & 1,484/2,174 & 1,556/2,247 & 1,561/2,261 & 1,565/2,241 \\
    \bottomrule
  \end{tabular}
  \label{tab:checkpoints-results}
\end{table}

We further analyze how each checkpoint contributes to the overall pool of correct patches. \Cref{tab:checkpoints-portion} shows the contribution of each checkpoint and the manual empty patch. Almost all the checkpoints contribute to the final results. The added manual deletion patch also has a positive contribution.

\begin{table}
  \caption{How many of the patches come from each checkpoint. Results are shown as \textit{correct/plausible}.}
  \centering
  \begin{tabular}{lcccccc}
    \toprule
    Benchmark         & manual  & checkpoint1 & checkpoint2 & checkpoint3 & checkpoint4 & checkpoint5 \\
    \midrule
    Defects4J (v1.2)  & 8/10    & 11/15       & 8/12        & 19/25       & 10/19       & 11/13       \\
    Defects4J (v2.0)  & 8/11    & 4/14        & 10/16       & 14/23       & 9/16        & 11/23       \\
    Bears             & 2/4     & 1/1         & 1/1         & 12/14       & 6/8         & 2/5         \\
    QuixBugs (Java)   & 1/1     & 9/10        & 5/5         & 6/6         & 4/4         & 0/0         \\
    QuixBugs (Python) & 4/5     & 5/5         & 4/4         & 8/8         & 8/8         & 0/0         \\
    Codeflaws         & 240/291 & 268/389     & 301/406     & 306/398     & 346/462     & 303/413     \\
    ManyBugs          & 2/-     & 2/-         & 2/-         & 3/-         & 3/-         & 3/-         \\
    BugAID            & 0/-     & 0/-         & 3/-         & 2/-         & 0/-         & 0/-         \\
    \midrule
    Total             & 265/322 & 300/434     & 334/444     & 370/474     & 386/517     & 330/454     \\
    \bottomrule
  \end{tabular}
  \label{tab:checkpoints-portion}
\end{table}

\begin{figure}
  \centering
  \includegraphics{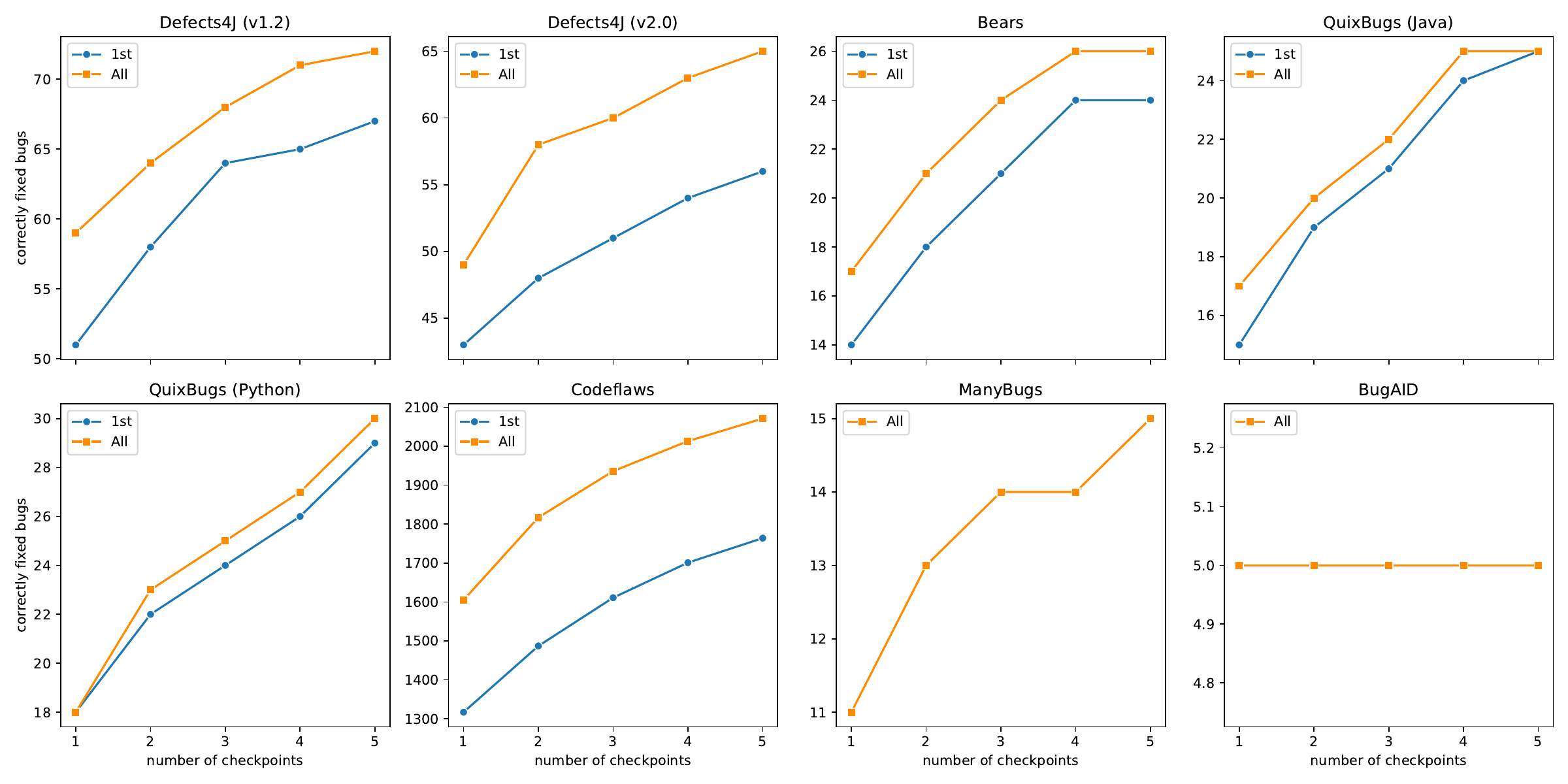}
  \caption{Results with incremental addition of checkpoints.}
  \label{fig:checkpoints-increment}
\end{figure}

\Cref{fig:checkpoints-increment} shows the result of incrementally adding the patches of each checkpoint to the generated patches of previous checkpoints. We can see that for most benchmarks, adding more checkpoints leads to better results both when considering the first plausible patch and all the plausible patches. However, for some benchmarks, such as BugAID, adding more checkpoints does not improve the results. This confirms the similar findings of \citet{lutellierCoCoNuTCombiningContextaware2020}.

\begin{tcolorbox}[title=RQ3 takeaways]

  The ensemble of checkpoints improves T5APR's performance over each checkpoint independently. The best-performing checkpoint varies for different benchmarks, suggesting that checkpoints complement each other. Almost all the checkpoints contribute to the final results. Adding more checkpoints generally improves results, but not always.

\end{tcolorbox}

\subsection{RQ4: Multilingual and monolingual}

In addition to the multilingual model, we also train models under the same setting as the multilingual model but only using training data of a single programming language. We then use monolingual models to generate patches for the benchmarks in the same language. \Cref{tab:multi-mono} shows the comparison of the first plausible correct patches of multilingual and monolingual models. The multilingual model outperforms the monolingual models for most benchmarks, except for ManyBugs and BugAID. Note that for ManyBugs and BugAID, running the validation step could change the results, and there might be correct patches that we have missed. Furthermore, the multilingual model fixes 426 unique bugs across all the benchmarks that the monolingual models do not fix, while the monolingual models fix 120 unique bugs. This highlights the benefit of leveraging multiple programming languages for training as it transfers bug patterns across languages.

\begin{table}
  \caption{Comparison of results of multilingual and monolingual models. Results are shown as \textit{correct/plausible}.}
  \centering
  \begin{tabular}{lcc}
    \toprule
    Benchmark         & Multilingual & Monolingual \\
    \midrule
    Defects4J (v1.2)  & 67/94        & 55/93       \\
    Defects4J (v2.0)  & 56/103       & 52/98       \\
    Bears             & 24/33        & 21/34       \\
    QuixBugs (Java)   & 25/26        & 21/24       \\
    QuixBugs (Python) & 29/30        & 26/28       \\
    Codeflaws         & 1,764/2,359  & 1,482/2,357 \\
    ManyBugs          & 15/-         & 16/-        \\
    BugAID            & 5/-          & 6/-         \\
    \midrule
    Total             & 1,985/2,645  & 1,679/2,634 \\
    \bottomrule
  \end{tabular}
  \label{tab:multi-mono}
\end{table}

\begin{tcolorbox}[title=RQ4 takeaways]

  T5APR's multilingual model outperforms the monolingual models on most benchmarks. The multilingual model fixes 426 unique bugs across all the benchmarks that the monolingual models do not fix. These results show the benefit of using multiple programming languages for training, which enables cross-lingual transfer learning.

\end{tcolorbox}

\subsection{Discussion}

Our results have several implications for both researchers and practitioners in the field of APR.
For researchers, our work provides a new perspective on the potential of leveraging pre-trained transformer models with multitask learning for multilingual program repair.
We show that even a relatively small model, when compared to typical large language model sizes, can handle multiple languages, eliminating the need for training separate models for each language. T5APR outperforms many existing approaches with just one epoch of fine-tuning, demonstrating that a substantial amount of resources is not a prerequisite for effective multilingual APR. Our work also highlights that although a single checkpoint may not represent all kinds of bugs, there is an efficient way to combat this by using the checkpoint ensemble strategy. Patch ranking remains a crucial aspect of APR research. Future studies should explore more effective ranking strategies to further improve the prioritization of correct patches, thereby reducing the time and effort required for developers to identify and fix bugs in their code. Multilingual training offers a promising direction for APR. By leveraging training data from multiple programming languages, models can potentially learn transferable bug-fixing patterns that generalize well across languages.

For practitioners, our work offers a practical and scalable solution for bug fixing across different languages and domains. Smaller models are easier and less expensive to deploy, and they can even be deployed on the client side to generate patches locally. APR approaches can integrate into continuous integration (CI) pipelines or development environments to be next to their primary source of debugging data, and with each commit, run the test suite and look for bugs to fix \citep{urliHowDesignProgram2018}. The ability to fix bugs in multiple languages makes T5APR particularly useful for developers working on multilingual codebases, which are increasingly common in modern software development. However, while T5APR offers promising results, the existence of plausible but incorrect patches shows that it is still essential for developers to validate the generated patches using their domain knowledge and expertise before applying them to production code.

\subsection{Threats to validity}

In this section, we outline possible threats that could impact the validity of our experimental findings and discuss how we mitigated them.

A major threat to internal validity is the potential fault in manual patch correctness assessment, which may result in misclassification or bias due to a lack of expertise or mistakes \citep{yeAutomatedPatchAssessment2021}. This is a common threat for all program repair results based on manual assessment \citep{yeComprehensiveStudyAutomatic2021}. To alleviate this threat, we compared our patches to patches generated by existing tools and carefully checked their semantic equivalency to reference developer patches. For results of other tools, we use the reported performance number in the paper of approaches and cross-checked them with patches in their repositories. Another threat to internal validity relates to potential fault in the implementation and hyperparameter configuration we used. We have double-checked our implementation, and to ensure reproducibility, we used fixed manual seed values wherever possible. To further mitigate these threats, we make all generated patches and our source code publicly available for verification and review by other researchers.

A third threat to internal validity comes from using CodeT5 as our base model, which is trained on large amounts of open-source code snippets. This means that its training data could overlap with our evaluation benchmarks. This issue is hard to address since retraining this model would require significant resources. However, some factors could mitigate this concern: First, the overlapping data, if any, would be a very small fraction of the training data. Second, both the correct and incorrect program versions would likely be present in the training data without any labels indicating which one is correct or incorrect since the pre-training objective of CodeT5 is different, and it was never specifically trained for the task of program repair. The same issue is present in approaches that use Codex or ChatGPT models \citep{prennerCanOpenAICodex2022,sobaniaAnalysisAutomaticBug2023}. Our APR training data is collected up to the date of bugs in our evaluation benchmark to avoid any overlap.

A threat to external validity is a threat that our approach might not be generalizable to fixing bugs outside the tested bugs benchmarks as shown by \citet{durieuxEmpiricalReviewJava2019} by the phenomenon of ``benchmark overfitting'' in program repair. We use six different benchmarks in four different programming languages with up to 5,257 real-world bugs to address this issue. Evaluation on more benchmarks (e.g., Bugs.jar \citep{sahaBugsJarLargeScale2018}, BugsJS \citep{gyimesiBugsJSBenchmarkJavaScript2019}, and BugsInPy \citep{widyasariBugsInPyDatabaseExisting2020}) could be done in the future.

\section{Related work}
\label{sec:related-work}

Automated program repair (APR) is a rapidly evolving and diverse field with a wide range of research and development efforts. In this section, we highlight some of the works that closely relate to our work while acknowledging that there are many other important contributions. For a more comprehensive survey of APR literature, we direct readers to recent surveys in the field \citep{gaoProgramRepair2022,huangSurveyAutomatedProgram2023,legouesAutomatedProgramRepair2019,monperrusLivingReviewAutomated2018}.

One well-established class of APR techniques is search-based methods that involve syntactic manipulation of code by applying different edit patterns. These techniques use a search algorithm to iteratively explore the space of possible code changes in order to find a plausible patch. Examples of such techniques include GenProg \citep{legouesGenProgGenericMethod2012}, SimFix \citep{jiangShapingProgramRepair2018}, and VarFix \citep{wongVarFixBalancingEdit2021}, among others.

Another class of APR approaches involves semantic analysis of code. These techniques use a set of constraint specifications to transform the program repair problem into a constraint solver problem and identify potential fixes that preserve the program's intended semantics. Examples of these techniques include SemFix \citep{nguyenSemFixProgramRepair2013}, Angelix \citep{mechtaevAngelixScalableMultiline2016}, and SOSRepair \citep{afzalSOSRepairExpressiveSemantic2019}.

Template-based methods generate repair patches by applying a predefined program fix template to the faulty code. A fix template specifies how to modify the code to fix a certain type of bug. Fix templates can be either manually defined \citep{liuYouCannotFix2019} or automatically mined from code repositories \citep{koyuncuFixMinerMiningRelevant2020,liuTBarRevisitingTemplatebased2019}.

Recently, learning-based techniques have gained traction in automatically fixing software bugs by learning from extensive code repositories. These techniques mostly employ neural networks to automatically generate correct code from buggy source code and are called neural program repair tools. These tools use a variety of sequence-to-sequence, neural machine translation (NMT), graph-to-sequence, and other deep learning models to generate patches as sequences of tokens \citep{chenSequenceRSequencetoSequenceLearning2019,jiangCURECodeAwareNeural2021,lutellierCoCoNuTCombiningContextaware2020,yeNeuralProgramRepair2022} or edits \citep{chakrabortyCODITCodeEditing2022,dingPatchingTranslationData2020,liDLFixContextbasedCode2020}. Thanks to the strong learning capabilities of these models, neural program repair techniques learn to capture the relationship between buggy and correct code, eliminating the need for manual design of fix patterns or feature templates and have outperformed many existing approaches. These works mostly use supervised learning on past bug-fixing commits. There are also works that use self-supervised learning that automatically generate training samples \citep{allamanisSelfSupervisedBugDetection2021,yasunagaGraphbasedSelfSupervisedProgram2020,yasunagaBreakItFixItUnsupervisedLearning2021,yeSelfAPRSelfsupervisedProgram2023}. Our work uses a sequence-to-sequence model and belongs to the supervised learning category.

\citet{tufanoEmpiricalStudyLearning2019} present an empirical study on using NMT to learn how to fix bugs in Java code. The authors mine a large dataset of bug-fixing commits from GitHub and extract method-level pairs of buggy and fixed code. They abstract the code to reduce the vocabulary size and train an encoder-decoder model to learn how to transform buggy code into fixed code.
\citet{chenSequenceRSequencetoSequenceLearning2019} introduce SequenceR, an end-to-end approach to program repair based on sequence-to-sequence learning on source code that sees the APR task as a translation from buggy to correct source code. A copy mechanism is used to handle the large and unlimited vocabulary of code, and an abstract buggy context is constructed to capture the relevant information for generating patches. The model is trained on a large dataset of one-line bug-fixing commits from GitHub and evaluated on CodRep \citep{chenCodRepMachineLearning2018} and Defects4J \citep{justDefects4JDatabaseExisting2014} benchmarks.
\citet{zhuSyntaxguidedEditDecoder2021} present Recoder, an approach for APR that uses a syntax-guided edit decoder with a provider/decider architecture to generate patches that are syntactically correct and context-aware. Recoder also introduces placeholder generation to handle project-specific identifiers and generates edits rather than modified code.
KNOD is proposed by \citet{jiangKNODDomainKnowledge2023}, which presents a tree decoder and a domain-rule distillation module. The tree decoder directly generates abstract syntax trees of patches in three stages: parent selection, edge generation, and node generation. The domain-rule distillation module enforces syntactic and semantic rules on the decoder during both training and inference phases.

\citet{yeNeuralProgramRepair2022} propose RewardRepair, a neural repair model that incorporates compilation and test execution information into the training objective. RewardRepair uses a discriminative model to reward patches that compile and pass test cases and penalize patches that are identical to the buggy code or introduce regression errors. The reward signal modulates the cross-entropy loss before backpropagation, guiding the neural network to generate high-quality and more compilable patches as shown in \cref{tab:cr-compare}. RewardRepair has two training phases, a syntactic training and a semantic one.

The work by \citet{jiangCURECodeAwareNeural2021} introduces CURE, a code-aware NMT technique for APR. CURE uses three techniques to improve the search space and the search strategy for generating patches: subword tokenization, pre-trained programming language model, and code-aware beam search to generate more compilable patches. CURE demonstrates the effectiveness of applying code awareness to NMT models for the APR task.
CURE differs from our approach in several aspects. CURE pre-trains its model exclusively using Java code and a standard causal language modeling task for pre-training a GPT model, while we use CodeT5, an encoder-decoder model that is pre-trained on multiple programming languages and more diverse pre-training tasks to better understand source code. Additionally, CURE's tokenizer is trained on their Java corpus, while ours is trained on the multilingual corpus of CodeT5. In contrast to CURE, we use the vanilla beam search to generate patches, which is simpler and faster than the code-aware one that CURE uses but may be less effective. However, beam search is an independent component, and we can incorporate the code-aware beam search into our tool in future work.

Most of these works target Java or are only evaluated on Java language benchmarks. Two works that are closest to our work and are evaluated on multiple programming languages are CoCoNuT by \citet{lutellierCoCoNuTCombiningContextaware2020} and CIRCLE by \citet{yuanCIRCLEContinualRepair2022}.

CoCoNuT is an APR technique that uses NMT to learn from bug fixes in open-source repositories. CoCoNuT has three main contributions: A context-aware NMT architecture that uses two separate encoders with fully convolutional layers (FConv) to represent the buggy line and its surrounding context; An ensemble approach that combines different models with different levels of complexity to capture various repair strategies; Cross-language portability that allows CoCoNuT to be applied to four programming languages (Java, Python, C, and JavaScript).
Our approach differs from this work in several ways. First, our approach can handle multiple programming languages with a unified model, unlike CoCoNuT, which requires individual models for each programming language. Second, we use a pre-trained programming language model that learns from a large software codebase to capture code syntax and developer-like coding style. Third, we use checkpoint ensemble for training efficiency, as opposed to CoCoNuT's model ensemble for each language. We combine these techniques to form a novel APR architecture using a text-to-text transformer model for patch generation across multiple programming languages.

\citet{yuanCIRCLEContinualRepair2022} propose CIRCLE, a method for APR that can handle multiple programming languages through continual learning. The method consists of five components: a prompt-based data representation, a T5-based model, a difficulty-based example replay, an elastic-based parameter updating regularization, and a re-repairing mechanism. The prompt-based data representation converts the bug-fixing task into a fill-in-the-blank task that is suitable for the pre-trained T5 model. The difficulty-based example replay and the elastic-based regularization are two continual learning strategies that prevent the model from catastrophic forgetting. The re-repairing mechanism is a post-processing step that corrects the errors caused by crossing languages.
The major differences between CIRCLE and our work are the following: We formulate APR as a multitask learning task, which is simpler and allows the trained model to remain relevant for a long time \citep{lutellierCoCoNuTCombiningContextaware2020}, so we do not need frequent retraining. By using a specific control prefix for each language, we do not need re-repairing of generated patches, and we get patches in the correct syntax of the target language. We also use a pre-trained code tokenizer and model instead of a pre-trained NLP tokenizer and model for better performance on APR tasks. Tokenizers that are trained on code usually generate fewer tokens for source code. Also, tokenizers that are only trained on natural text often fail to handle code tokens well. This is why CIRCLE needs the re-repairing step to fix unknown tokens of the generated patches.

Several studies have explored the potential of large language models like OpenAI's Codex and ChatGPT for APR. \citet{prennerCanOpenAICodex2022} evaluate the performance of Codex on the QuixBugs benchmark. Codex is a GPT-3-like model that can generate code snippets from natural language descriptions or partial code inputs. The authors experiment with different prompts to trigger Codex's bug-fixing ability, such as providing hints, docstrings, or input-output examples. They find that Codex is competitive with state-of-the-art neural repair techniques, especially for Python, and that the choice of prompt has a significant impact on the repair success.
Similarly, \citet{sobaniaAnalysisAutomaticBug2023} assess the automatic bug fixing performance of ChatGPT. They compare ChatGPT with Codex, CoCoNuT, and several standard APR approaches on the QuixBugs benchmark. The study finds that ChatGPT has a similar performance to Codex and CoCoNuT and outperforms the standard APR approaches in terms of the number of fixed bugs. The authors also analyze the types of responses generated by ChatGPT and show that providing hints to ChatGPT can improve its success rate.

\begin{table}
  \caption{Comparison of different learning-based APR approaches.}
  \centering
  \scriptsize
  \begin{tabular}{lccll}
    \toprule
    Approach                                                      & Evaluated Language          & Beam size & Tokenizer               & Model                                  \\
    \midrule
    \citet{tufanoEmpiricalStudyLearning2019}                      & Java                        & 50        & Word                    & Encoder-decoder LSTM                   \\
    SequenceR \citep{chenSequenceRSequencetoSequenceLearning2019} & Java                        & 50        & Word                    & Encoder-decoder LSTM                   \\
    DLFix \citep{liDLFixContextbasedCode2020}                     & Java                        & -         & Word                    & Tree-based LSTM                        \\
    CoCoNuT \citep{lutellierCoCoNuTCombiningContextaware2020}     & Java, Python, C, JavaScript & 1,000     & Word                    & FConv                                  \\
    CURE \citep{jiangCURECodeAwareNeural2021}                     & Java                        & 1,000     & Subword (BPE)           & GPT + FConv                            \\
    Recoder \citep{zhuSyntaxguidedEditDecoder2021}                & Java                        & 100       & Word                    & Tree-based transformer                 \\
    CIRCLE \citep{yuanCIRCLEContinualRepair2022}                  & Java, Python, C, JavaScript & 250       & Subword (SentencePiece) & T5                                     \\
    RewardRepair \citep{yeNeuralProgramRepair2022}                & Java                        & 200       & Subword (SentencePiece) & Transformer                            \\
    Codex \citep{prennerCanOpenAICodex2022}                       & Java, Python                & -         & Subword (BPE)           & Davinci-codex                          \\
    ChatGPT \citep{sobaniaAnalysisAutomaticBug2023}               & Java                        & -         & Subword (BPE)           & GPT-3.5-turbo                          \\
    KNOD \citep{jiangKNODDomainKnowledge2023}                     & Java                        & 1,000     & Word                    & Graph transformer + Tree-based decoder \\
    \midrule
    T5APR                                                         & Java, Python, C, JavaScript & 100       & Subword (BPE)           & CodeT5                                 \\
    \bottomrule
  \end{tabular}
  \label{tab:related-work}
\end{table}

\Cref{tab:related-work} presents a comparison of learning-based APR approaches. It summarizes the programming languages each approach is evaluated on, the beam size used in the model's search strategy, the type of tokenizer, and the underlying model architecture.

Although most of the APR techniques focus on single-hunk bugs, the challenge of addressing multi-hunk bugs has attracted attention as well \citep{liDEARNovelDeep2022,sahaHarnessingEvolutionMultiHunk2019,wongVarFixBalancingEdit2021,yeITERIterativeNeural2024}. T5APR also targets a limited subset of multi-hunk bugs with the potential for further expansion in future work.

There are also multilingual repair techniques that address compilation and syntax issues \citep{joshiRepairNearlyGeneration2023,yasunagaBreakItFixItUnsupervisedLearning2021}. However, these techniques focus on a different problem. We target dynamic and functional errors, which can persist even when programs compile successfully, leading to incorrect behavior.

\section{Conclusion}
\label{sec:conclusion}

In this paper, we proposed T5APR, a novel approach for automated program repair (APR) that leverages the power of the CodeT5 text-to-text transformer model. Our method addresses program repair challenges across various programming languages, offering a unified solution for bug fixing. Our approach has several noteworthy contributions. We demonstrated the ability of T5APR to efficiently handle multiple programming languages, fixing bugs in Java, Python, C, and JavaScript. The checkpoint ensemble strategy further improves the reliability and performance of our approach, providing more robust patches for different bugs.

We conducted an extensive experimental evaluation to highlight the effectiveness, generalizability, and competitiveness of T5APR. T5APR correctly fixed 1,985 bugs out of 5,257 bugs across six benchmarks, including 1,442 bugs that other compared state-of-the-art repair techniques did not fix. Moreover, the patch ranking comparison showed the promising performance of T5APR in terms of generating high-ranking patches.

In addition to the contributions of this research, there are several directions for future work that can further enrich the field of APR. We can investigate the selection of context windows and the impact of using larger context windows beyond the immediate buggy context to find the optimal balance between information and computational efficiency. We can use a more advanced training and checkpoint selection process to further enhance T5APR's performance. One promising avenue is the exploration of low-rank adaptation (LoRA) \citep{huLoRALowRankAdaptation2021} for efficient fine-tuning of large language models, such as Mistral 7B \citep{jiangMistral7B2023}.
Additionally, we can extend our approach to handle more complex scenarios, such as multi-hunk bugs with different changes in multiple locations. We can also further expand the supported languages and evaluation benchmarks, even languages that were not part of the pre-training of CodeT5 but have similar syntax and semantics, as multilingual learning especially helps knowledge transfer in low-resource languages \citep{zugnerLanguageAgnosticRepresentationLearning2020}. As the field progresses, the development of explainable patch generation techniques \citep{liangHowExplainPatch2019} and close collaboration with software developers could foster the usability and trustworthiness of automated repair solutions.

\bibliographystyle{plainnat}
\bibliography{references}  


\end{document}